\pdfoutput=1

\documentclass[11pt,twoside,a4paper,cmspaper,final,collab]{cms-tdr}
\def\svnVersion{494048P}\def\cmsCernNoTag{CERN-EP-2018-326}\def\cmsCernDate{\today}\def\cmsMessage{Published in the Journal of High Energy Physics as \href{http://dx.doi.org/10.1007/JHEP04(2019)114}{\doi{10.1007/JHEP04(2019)114.}}}

\begin{document}\cmsNoteHeader{EXO-17-025}

\hyphenation{had-ron-i-za-tion}
\hyphenation{cal-or-i-me-ter}
\hyphenation{de-vices}
\RCS$HeadURL: svn+ssh://svn.cern.ch/reps/tdr2/papers/EXO-17-025/trunk/EXO-17-025.tex $
\RCS$Id: EXO-17-025.tex 494044 2019-04-22 17:42:17Z sturdy $

\newlength\cmsFigWidth
\newlength\cmsTabSkip\setlength{\cmsTabSkip}{1ex}
\providecommand{\cmsTable}[1]{\resizebox{\textwidth}{!}{#1}}
\providecommand{\NA}{\ensuremath{\text{---}}}
\providecommand{\CL}{CL\xspace}

\newcommand{\massADD}[1]{\ensuremath{M_{\mathrm{#1}}}\xspace}
\newcommand{\lambdaCIADD}[1]{\ensuremath{\Lambda_{\mathrm{#1}}}\xspace}
\newcommand{\etaCIADD}[1]{\ensuremath{\eta_{\mathrm{#1}}}\xspace}
\newcommand{\eeLumi}{35.9}
\newcommand{\uuLumi}{36.3}
\newcommand{\mll}{\ensuremath{m_{\ell\ell}}\xspace}
\newcommand{\mee}{\ensuremath{m_{\Pe\Pe}}\xspace}
\newcommand{\muu}{\ensuremath{m_{\PGm\PGm}}\xspace}
\newcommand{\mgg}{\ensuremath{m_{\Pgg\Pgg}}\xspace}

\cmsNoteHeader{EXO-17-025}
\title{Search for contact interactions and large extra dimensions in the dilepton mass spectra from proton-proton collisions at $\sqrt{s}=13\TeV$}

\date{\today}

\abstract{
A search for nonresonant excesses in the invariant mass spectra of electron and muon pairs is presented.
The analysis is based on data from proton-proton collisions at a center-of-mass energy of 13\TeV recorded by the CMS experiment in 2016, corresponding to a total integrated luminosity of 36\fbinv.
No significant deviation from the standard model is observed.
Limits are set at 95\% confidence level on energy scales for two general classes of nonresonant models.
For a class of fermion contact interaction models, lower limits ranging from 20 to 32\TeV are set on the characteristic compositeness scale $\Lambda$.
For the Arkani-Hamed, Dimopoulos, and Dvali model of large extra dimensions, the first results in the dilepton final state at 13\TeV are reported, and values of the ultraviolet cutoff parameter \lambdaCIADD{T} below 6.9\TeV are excluded.
A combination with recent CMS diphoton results improves this exclusion to \lambdaCIADD{T} below 7.7\TeV, providing the most sensitive limits to date in nonhadronic final states.
}

\hypersetup{%
pdfauthor={CMS Collaboration},%
pdftitle={Search for contact interactions and large extra dimensions in the dilepton mass spectra from proton-proton collisions at sqrt(s) = 13 TeV},%
pdfsubject={CMS},%
pdfkeywords={CMS, Physics, Exotica, Dimuon, Dielectron, ADD, LED, Large Extra Dimensions, Extra Dimensions, Gravitons, CI, Contact Interactions}
}

\maketitle

\section{Introduction}
\label{sec:introduction}
Nonresonant enhancements of the production rate of high invariant mass lepton pairs in proton-proton (\Pp\Pp) collisions have been predicted in several models~\cite{Eichten:1984eu,arkani98:hlz} of phenomena beyond the standard model (SM).
In these models, the differential cross section for the production of charged lepton pairs can be described by the equation:
\begin{linenomath}
\begin{equation}
  \dd{\sigma_{\text{X}\to\ell\ell}}{\mll}=
  \dd{\sigma_{\text{DY}}}{\mll}
  +\etaCIADD{X}\mathcal{I}(\mll)
  +\etaCIADD{X}^{2}\mathcal{S}(\mll)\text{,}
  \label{eq:dilepxs}
\end{equation}
\end{linenomath}
where \mll is the invariant mass of the two leptons, $\rd{\sigma_{\text{DY}}}/\rd{\mll}$ is the SM Drell--Yan (DY) differential cross section, \etaCIADD{X} is a model specific form factor, and the signal contribution terms are separated into an interference term ($\mathcal{I}$) and a pure signal term ($\mathcal{S}$).
Interference between new physical processes and the SM DY process is possible when the new process acts on the same initial state and yields the same final state.
For the analysis presented in this paper we consider two nonresonant scenarios: a contact interaction arising from the existence of fermion substructure; and the effects of virtual spin-2 gravitons as predicted by models with large extra dimensions.

The existence of three generations of quarks and leptons has led to speculation~\cite{Eichten:1984eu} that these particles may be composed of more fundamental constituents, which have been called ``preons''.
The preons would account for the properties of quarks and leptons via a new strong gauge interaction, analogous to the color interaction in quantum chromodynamics (QCD).
Below a given energy scale $\Lambda$, the main effect of this QCD-like interaction is to bind the preons into singlet states with respect to the new gauge interaction.
Given the present limits on the substructure of quarks and leptons, it is expected that $\Lambda$ would be on the order of at least several TeV.
For parton interactions at a center-of-mass energy $\sqrt{\hat{s}}$ much lower than $\Lambda$, the presence of preon bound states would result in a flavor-diagonal ``contact interaction'' (CI)~\cite{Eichten:1983hw}.
Assuming quarks and leptons share common constituents, the Lagrangian for the CI process $\cPq\cPaq\to\ell\ell$, where $\ell$ is a charged lepton, can be expressed as
\begin{linenomath}
\begin{equation}
  \mathcal{L}_{\cPq\ell}=\frac{g^2_{\text{contact}}}{\Lambda^2}
  \left[
    \begin{aligned}
        &\etaCIADD{LL}(\cPaq_\text{L}\gamma^\mu\cPq_\text{L})(\overline{\ell}_\text{L}\gamma_{\mu}\ell_\text{L})
      + \etaCIADD{RR}(\cPaq_\text{R}\gamma^\mu\cPq_\text{R})(\overline{\ell}_\text{R}\gamma_{\mu}\ell_\text{R}) \\ %
      + &\etaCIADD{LR}(\cPaq_\text{L}\gamma^\mu\cPq_\text{L})(\overline{\ell}_\text{R}\gamma_{\mu}\ell_\text{R})
      + \etaCIADD{RL}(\cPaq_\text{R}\gamma^\mu\cPq_\text{R})(\overline{\ell}_\text{L}\gamma_{\mu}\ell_\text{L})
    \end{aligned}
  \right],
  \label{eq:cil}
\end{equation}
\end{linenomath}
where $\cPq_\text{L}=(\cPqu,\cPqd)_\text{L}$ is a left-handed quark doublet;
$\cPq_\text{R}$ represents a sum over the right-handed quark singlets (\cPqu- and \cPqd-type);
and $\ell_\text{L}$ and $\ell_\text{R}$ are the left- and right-handed leptons, respectively.
By convention, $g_{\text{contact}}^2/{4\pi}=1$ and the helicity parameters \etaCIADD{ij} are taken to have unit magnitude.
The compositeness scale, represented by $\Lambda$, is potentially different for each of the individual terms in the Lagrangian.
Therefore, the individual helicity currents for ``left-left'' ($\text{LL}$), ``right-right'' ($\text{RR}$), and the combination of ``left-right'' ($\text{LR}$) and ``right-left'' ($\text{RL}$) in Eq.~(\ref{eq:cil}), together with their scales (\lambdaCIADD{LL}, \lambdaCIADD{RR}, and \lambdaCIADD{LR}), are considered separately in this search, and in each case all other currents are assumed to be zero.
The combination of $\text{LR}$ and $\text{RL}$ is referred to simply as $\text{LR}$ throughout the paper.
A given \etaCIADD{ij} can be related to the form factor in the differential cross section in Eq.~(\ref{eq:dilepxs}) by
\begin{linenomath}
\begin{equation}
  \etaCIADD{X} = -\frac{\etaCIADD{ij}}{\lambdaCIADD{ij}^{2}},
  \label{eq:ciprefact}
\end{equation}
\end{linenomath}
where both constructive ($\etaCIADD{ij}<0$) and destructive ($\etaCIADD{ij}>0$) interference with DY processes are possible.

Theories extending the SM with additional dimensions have been studied extensively~\cite{PDG2018}.
The model with large extra dimensions developed by Arkani-Hamed, Dimopoulos, and Dvali (ADD)~\cite{arkani98:hlz} describes quantum gravity as an effective field theory.
It has the potential to solve, at the TeV scale, the so-called ``hierarchy problem'', which arises from the large difference between the Higgs boson mass~\cite{PDG2018} and the energy scale, referred to as the Planck mass \massADD{Pl}, at which gravity is expected to become strong.
This is achieved via an extension of spacetime by $n$ additional compactified spatial dimensions of size $L$.
In the ADD model, all SM particles are confined to the four-dimensional subspace (the brane), while gravity can propagate to all $D=n+4$ dimensions (the bulk).
If $L$ is sufficiently large, the $D$-dimensional fundamental Planck mass \massADD{D}, which is related to \massADD{Pl} in three dimensions by
\begin{linenomath}
\begin{equation}
  \massADD{D}^{2+n} = \massADD{Pl}^2/L^n,
  \label{eq:led-planck}
\end{equation}
\end{linenomath}
can then be probed at the \TeV scale.
The aforementioned compactification of the additional dimensions results in periodic boundary conditions, and thus a quasi-continuous spectrum of Kaluza--Klein graviton modes.
As the interaction scale increases, more graviton modes are excited, leading the ADD model to predict a nonresonant excess of lepton pairs at high dilepton masses originating from the decay of virtual gravitons.
These processes can be characterized by the single energy cutoff scale \lambdaCIADD{T} in the Giudice--Rattazzi--Wells (GRW) convention \cite{GRW99:extradim}, the string scale \massADD{S} in the Hewett convention~\cite{Hewett:1998sn}, or the number of additional dimensions $n$ in conjunction with \massADD{S} in the Han--Lykken--Zhang (HLZ) convention~\cite{han99:hlz}.
The generic form factor \etaCIADD{X} is replaced by \etaCIADD{G} in Eq.~(\ref{eq:dilepxs}), which depends on the chosen convention:
\begin{linenomath}
\begin{align}
  \label{eq:grw}
  & \text{GRW:}    & \etaCIADD{G} &= \frac{1}{\lambdaCIADD{T}^4}; \\
  \label{eq:hew}
  & \text{Hewett:} & \etaCIADD{G} &= \frac{2}{\pi} \frac{\lambda}{\massADD{S}^4} \quad\text{with}\,\lambda=\pm1; \\
  \label{eq:hlz}
  & \text{HLZ:}    & \etaCIADD{G} &=
  \begin{cases}
    \ln\left({\massADD{S}^2}/{\hat{s}}\right)\frac{1}{\massADD{S}^4} & \text{for}\,n = 2        \\
    \frac{2}{n-2} \frac{1}{\massADD{S}^4}                            & \text{for}\,n > 2.
  \end{cases}
\end{align}
\end{linenomath}
Of the three, only the Hewett convention allows both constructive and destructive interference with the SM DY process, but in this paper only the constructive case ($\Lambda=+1$) is considered.
Relative to CI models, interference with DY in the ADD model is more limited as the production of virtual gravitons is dominated by gluon-induced processes.
Both \lambdaCIADD{T} and \massADD{S} function as ultraviolet (UV) cutoff parameters, indicating the energy scale up to which the effective field theory provides reliable predictions.
Beyond this point, a description of quantum gravity becomes necessary to accurately describe particle interactions.

The analysis presented in this paper focuses on dilepton (electron or muon) events produced in {\Pp\Pp} collisions at a center-of-mass energy of 13\TeV at the CERN LHC.
The data sample was recorded by the CMS experiment in 2016, and corresponds to an integrated luminosity of {\eeLumi} ({\uuLumi})\fbinv for the electron (muon) channel.

For both the CI and ADD models, this paper extends previous results from CMS at 8\TeV~\cite{Khachatryan:2014fba}, and complements the recent CMS search at 13\TeV for resonant phenomena~\cite{Sirunyan:2018exx} in dilepton final states.
Additional constraints on these models from diphoton and dijet final states have been reported by CMS~\cite{Sirunyan:2018wnk,Sirunyan:2018wcm}.
The ATLAS Collaboration has presented similar results for these models in the dilepton final state, the most recent using data at 8\TeV~\cite{Aad:2014wca} for the ADD model and at 13\TeV~\cite{Aaboud:2017buh} for the CI model.

\section{The CMS detector}
\label{sec:detector}
The central feature of the CMS detector is a superconducting solenoid providing an axial magnetic field of 3.8\unit{T} and enclosing a silicon strip and pixel tracker, an electromagnetic calorimeter (ECAL), and a hadron calorimeter (HCAL).
The silicon tracker measures charged particles within the pseudorapidity range $\abs{\eta}<2.5$.
The ECAL and HCAL, each composed of a barrel and two endcap sections, extend over the range $\abs{\eta}<3$, while a forward calorimeter encompasses $3<\abs{\eta}<5$.

The muon detection system covers $\abs{\eta}<2.4$ with up to four layers of gas-ionization chambers installed outside the solenoid and sandwiched between the layers of the steel flux-return yoke.
Additional detectors and upgrades of electronics were installed before the beginning of the 13\TeV data collection period in 2015, yielding improved reconstruction performance for muons relative to the 8\TeV data collection period in 2012.
A more detailed description of the CMS detector, together with a definition of the coordinate system used and the relevant kinematic variables, can be found in Ref.~\cite{CMS-JINST}.

The CMS experiment has a two-level trigger system~\cite{Khachatryan:2016bia}.
The level-1 (L1) trigger, composed of custom hardware processors, selects events of interest using information from the calorimeters and muon detectors;
the software based high-level trigger (HLT) then uses the full event information, including that from the inner tracker, to select the events that are recorded for analysis.

\section{Lepton reconstruction and event selection}
\label{sec:objects}
A detailed description of the reconstruction and selection of electron and muon pairs used in this analysis can be found in Ref.~\cite{Khachatryan:2016zqb} and is briefly summarized below.

Candidate events in the electron channel are selected first by the L1 trigger, which requires two energy deposits (clusters) in the ECAL with transverse momentum $\pt>24\,(17)\GeV$, respectively.
A suite of L1 trigger algorithms, requiring single, highly energetic calorimeter clusters, has also been used to select events for this analysis to guard against potential inefficiencies of the primary trigger.
The HLT then requires that both electron candidates have $\pt>33\GeV$ and pass loose identification criteria.

Electron candidates are reconstructed by matching tracks originating from the nominal interaction point with ECAL energy clusters.
These clusters include the energy coming from bremsstrahlung photons.
The electron candidates are required to have $\pt>35\GeV$ and cluster pseudorapidity $\abs{\etaCIADD{C}}<1.44$ (barrel) or $1.57<\abs{\etaCIADD{C}}<2.50$ (endcap).
The intermediate region is excluded because of the reduced reconstruction quality of clusters in the overlap of the barrel and endcap components of the ECAL.

Furthermore, the candidates are required to pass a specialized selection, optimized for high-energy electrons~\cite{Khachatryan:2015hwa}, ensuring that the electron track is well reconstructed, that the transverse size of the ECAL cluster is consistent with that of an electron, and that there is minimal energy leakage into the HCAL.
Additionally, the electron candidate must be well isolated in the calorimeter and the tracker, within a cone of radius $\Delta R=\sqrt{\smash[b]{(\Delta\eta)^2+(\Delta\phi)^2}}=0.3$, where $\phi$ is the azimuthal angle.

For events in which two or more electrons meet all of the aforementioned requirements, all possible electron pair candidates are created.
For each of the pair candidates, at least one of the electrons is required to be in the barrel region.
Should more than one pair pass the selection, the pair with the largest \pt sum is used.

In the muon channel, events are selected by the L1 trigger requiring two muons, at least one of which must have transverse momentum $\pt>22\GeV$.
The HLT requires that at least one of the muons have $\abs{\eta}<2.4$ and $\pt>50\GeV$.
A separate HLT algorithm, with a threshold of $\pt>27\GeV$, is used to select a large event sample at the {\cPZ} boson peak ($60<\muu<120\GeV$), which is used to derive the normalization of the simulated backgrounds.

Muon candidates are required to have matching segments in the tracker and the muon system.
Further selection requirements are applied offline~\cite{Khachatryan:2014fba}, among which are the requirements that muon candidates must have $\abs{\eta}<2.4$ and $\pt>53\GeV$.
Isolated muon candidates are selected by requiring that the scalar sum of the transverse momenta of all tracks within a cone of $\Delta R<0.3$ around the muon must be less than 10\% of the muon \pt.
A dedicated algorithm~\cite{Sirunyan:2018fpa} is used for the reconstruction of muons with $\pt>200\GeV$, which accounts for radiative energy losses due to interactions of the highly energetic muons with the detector material.

Muon pairs are formed from oppositely charged muons, with one of the muons required to match the muon that triggered the event.
A $\chi^2$ fitting method is used to ensure that the muon candidate tracks are compatible with originating from a common vertex.
The three-dimensional angle between the two muon candidates is required to be less than $\pi-0.02$, to suppress muons originating from cosmic rays.
If more than one pair of muons pass all aforementioned requirements, the pair with the highest \pt sum is chosen.

The search region ($\mll>400\GeV$) is divided into two categories, depending on the location of the two leptons.
Events where both leptons are in the barrel region are called barrel-barrel (BB), while events where at least one lepton is in the endcap are called barrel-endcap (BE).
For the electron channel, events where both electrons are in the endcap region are ignored.
The efficiency to trigger, reconstruct, and select a lepton pair with invariant mass around 1\TeV is 69 (65)\% in the electron channel for BB (BE) events, while it is about 93\% for events in the muon channel.

\section{Background and signal estimation}
\label{sec:sigbkgdestimation}
The primary SM production channel for lepton pairs in this analysis is the DY process.
It is simulated with \POWHEG~\textsc{v2}~\cite{Nason:2004rx,Frixione:2007vw,Alioli:2010xd,Alioli:2008gx,Frixione:2007nw,Re:2010bp} at next-to-leading-order (NLO) in perturbative QCD, using the \textsc{nnpdf}~{3.0}~\cite{Ball:2014uwa} set of parton distribution functions (PDFs) and \PYTHIA~{8.205}~\cite{Sjostrand:2014zea} for parton showering and hadronization.
A mass-dependent correction factor is applied in order to reach next-to-next-to-leading order (NNLO) accuracy in perturbative QCD, and to account for weak effects at NLO, as well as pure quantum electrodynamics effects.
This factor is derived as the ratio of the cross sections calculated by \FEWZ 3.1b2~\cite{Li:2012wna} to those calculated with \POWHEG, using a combination of PDFs from \textsc{pdf4lhc15}~\cite{Botje:2011sn,Alekhin:2011sk,Butterworth:2015oua} and the \textsc{lux}~\cite{Manohar:2016nzj} PDF set for the photon PDFs.
This correction factor also accounts for photon-induced processes~\cite{Bourilkov:2016qum,Bourilkov:2016oet}, stemming from {\cPgg\cPgg} initial states.
The effect of these processes does not exceed 5\% for masses up to 2\TeV and reaches 15--20\% above 5\TeV~\cite{Bourilkov:2016oet}.
The simulation of the detector response is performed by \GEANTfour~\cite{Agostinelli:2002hh}.

Other background processes yielding lepton pairs in the signal region are the production of top quark pairs, single top quarks via {\PW\cPqt} production, and production of {\PW} boson pairs (\PW\PW).
These processes are simulated with \POWHEG~\cite{Nason:2004rx,Frixione:2007vw,Alioli:2010xd,Alioli:2008gx,Frixione:2007nw,Re:2010bp}, using \textsc{nnpdf}~{3.0} as the PDF set and a mix of \PYTHIA~{8.205 and 8.212} for showering and hadronization.
The top quark pair production cross section is calculated up to NNLO, including leading-log effects for soft gluon resummation, with \textsc{Top++}~{2.0}~\cite{Czakon:2011xx}, while the {\PW\cPqt} cross section has been calculated up to next-to-next-to-leading log accuracy~\cite{Kidonakis:2010ux}.
Cross sections for other processes have been calculated up to NNLO with \MCFM~{6.6}~\cite{Boughezal:2016wmq,Campbell:2015qma,Campbell:2011bn,Campbell:1999ah}.

In addition to the {\PW\PW} background produced with \POWHEG, {\PW\cPZ} and {\cPZ\cPZ} production is simulated inclusively at leading order (LO) with \PYTHIA, using the \textsc{nnpdf}~{2.3}~\cite{Ball:2012cx} PDF set.
Production of {\Pgt} lepton pairs through the DY process, which then decay to electron or muon pairs, is simulated at NLO with \MGvATNLO~{2.2.2}~\cite{Alwall:2014hca}, using the \textsc{nnpdf}~{3.0} PDF set and \PYTHIA for showering and hadronization.

The overall yield from these processes is then normalized to the data in the control region around the {\cPZ} boson peak.
Background from events containing jets that are misreconstructed as isolated leptons, is estimated from data using event samples enriched in QCD multijet events, as described in Ref.~\cite{Khachatryan:2014fba}.
The contribution of this background to the overall event sample is between 1--3\%.

{\tolerance=800 Each signal model, including interference with the DY process, is simulated at LO using \textsc{nnpdf}~{2.3} and \PYTHIA~{8.212 and 8.205} for the CI and ADD samples, respectively.
A dedicated \PYTHIA  DY sample is produced with the same generator settings and subtracted from the signal samples to obtain the respective signal yields.
No higher-order correction factor is applied to the signal samples of the CI model; for the ADD model, a mass-independent NLO correction factor of 1.3 is used.
While NNLO QCD predictions show that this correction factor can be as large as 1.6~\cite{Ahmed:2017}, and that it always exceeds 1.3 in the considered dilepton mass range, NLO electroweak corrections are not taken into account.
This motivates choosing the conservative value of 1.3, which also allows a direct comparison to previous results~\cite{Khachatryan:2014fba}.
\par}

To account for the effects of additional {\Pp\Pp} interactions within the same or nearby bunch crossings (``pileup''), additional minimum bias events are overlaid on the simulated events.
The simulated events are scaled to match the recorded luminosity, using the cross sections obtained as described above, and then reweighted so that their pileup distribution matches the one observed in the data.

\section{Systematic uncertainties}
\label{sec:uncertainties}
A summary of the systematic uncertainties in the SM background estimates is found in Table~\ref{tab:systematics-overview}, and brief descriptions of their determination are given below.
For each source, the corresponding relative uncertainty in the event yield is given separately for the electron and muon channels.
To illustrate the mass-dependent nature of some of the uncertainties, values are shown for two different invariant mass thresholds.
All of the mass-dependent uncertainties listed in Table~\ref{tab:systematics-overview} affect both the total number of events and the shape of the invariant mass distribution.

The efficiency of triggering, reconstructing, and selecting electrons is measured in simulated DY events and validated using data at the {\cPZ} boson peak.
The uncertainty in the electron energy scale of 2 (1)\% in the barrel (endcap) region has been used to derive the resulting uncertainty in the event yield.

The efficiency of the single-muon trigger to identify either of the two muons in the event has been measured using a sample of {\cPZ} boson candidate events, and is found to be independent of mass.
Uncertainty in the reconstruction and selection efficiency for muons leads to a corresponding uncertainty in event yield.
The uncertainty in muon efficiency, as a function of \pt and $\eta$, is determined from differences between data and simulation.
Because a potential bias in the muon \pt measurement may result in a bias in the dimuon mass scale, the muon curvature ($q/\pt$, where $q$ is the electric charge of the muon) distribution in data is compared to that obtained from simulation for different $\eta$ and $\phi$ ranges.
The measured bias is consistent with zero, and, along with the corresponding uncertainty, is propagated to the dimuon mass to derive the uncertainty in the event yield.
The muon \pt resolution and its uncertainty are determined using muons from events with Lorentz-boosted {\cPZ} bosons.
The uncertainty in the resolution is found to scale with \pt.

The remaining uncertainties are applicable to both the electron and muon channels.
The simulated backgrounds are normalized using data at the {\cPZ} boson peak, and a systematic uncertainty is assigned to cover the observed difference between data and simulation before normalization.
The uncertainty in the cross section calculation of the simulated diboson and {\ttbar} events is found to be a constant 7\%.
Uncertainty in the PDF leads to uncertainties in the simulated DY yields.
The uncertainty is determined with the \textsc{pdf4lhc} procedure~\cite{Botje:2011sn,Alekhin:2011sk,Butterworth:2015oua} using replicas of the \textsc{nnpdf}~{3.0} PDF set~\cite{Ball:2014uwa}.
Other uncertainties in the NNLO DY cross section, such as due to the scale of the strong coupling constant \alpS, have a negligible effect on the event yields.
The precision in estimating the misreconstructed jet background is limited by the amount of data at high dilepton mass, and a conservative uncertainty of 50\% is assigned.
The systematic uncertainty in the simulation of pileup is derived from the 5\% precision on the total inelastic {\Pp\Pp} scattering cross section that is used in the procedure to reweight the simulated event samples.
The cross section is varied by this uncertainty and used to reweight the simulated events, resulting in a variation in the invariant mass distribution for all simulated processes.

\begin{table}[!h]
  \centering
  \topcaption{
    Systematic uncertainties in the predicted SM yields for the electron and the muon channels, for two dilepton mass thresholds.
    Where noted, uncertainties are provided separately for events where both leptons are in the barrel region (BB), or where at least one of the leptons is in the endcap region (BE).
    Uncertainties that are mass-dependent affect both the event yield and the shape of the invariant mass distribution.
    The systematic uncertainties in the signal yields are largely the same as for the background, with a few exceptions as discussed in the text.
  }
  \cmsTable{
    \begin{tabular}{lcccc}
      \hline
      & \multicolumn{2}{c}{Electrons} & \multicolumn{2}{c}{Muons} \\
      \textbf{Uncertainty}            & $\mee>2\TeV$ & $\mee>4\TeV$ & $\muu>2\TeV$ & $\muu> 4\TeV$ \\
      \hline
      Electron trigger + selection efficiency BB (BE)  & \multicolumn{2}{c}{6 (8)\%}      & \NA           & \NA             \\
      Electron energy scale BB (BE)                    & 12.0 (6.7)\%    & 21.7 (11.0)\%  & \NA           & \NA             \\
      Muon trigger efficiency BB (BE)                  & \NA             & \NA            & \multicolumn{2}{c}{0.3 (0.7)\%} \\
      Muon ID efficiency BB (BE)                       & \NA             & \NA            & 0.8 (4.6)\%   & 1.7 (7.6)\%     \\
      Muon \pt resolution BB (BE)                      & \NA             & \NA            & 0.8 (1.4)\%   & 1.5 (2.3)\%     \\
      Muon \pt scale BB (BE)                           & \NA             & \NA            & 0.8 (2.8)\%   & 4.1 (12.1)\%    \\
      {\ttbar}/diboson cross section                   & \multicolumn{2}{c}{7\%}          & \multicolumn{2}{c}{7\%}         \\
      {\cPZ} boson peak normalization                  & \multicolumn{2}{c}{1\%}          & \multicolumn{2}{c}{5\%}         \\
      PDF                                              & 5.7\%           & 17.1\%         & 5.7\%         & 17.1\%          \\
      Multijet BB (BE)                                 & 0.1 (1.3)\%     & 0.1 (0.1)\%    & ${<}0.1\,(4.8)\%$ & ${<}0.1\,({<}0.1)\%$ \\
      Pileup reweighting BB (BE)                       & 0.5 (0.7)\%     & 0.4 (0.7)\%    & 0.2 (0.1)\%   & 0.2 (0.2)\%     \\
      MC statistics BB (BE)                            & 1.0 (1.8)\%     & 0.7 (1.7)\%    & 1.1 (1.3)\%   & 1.0 (2.0)\%     \\
      \hline
    \end{tabular}
  }
  \label{tab:systematics-overview}
\end{table}

The systematic uncertainties in the signal yields are largely the same as for the background, with a few exceptions.
The signal samples are normalized to the total integrated luminosity, rather than to the data at the {\cPZ} boson peak, and the uncertainty on the luminosity measurement is 2.5\%~\cite{CMS-PAS-LUM-17-001}.
Additionally, the uncertainties due to the cross sections and jet background estimation do not apply to the simulated signal events.

\section{Mass spectra and statistical analysis}
\label{sec:spectra}
\begin{figure}[!htb]
  \centering
  \includegraphics[width=0.5\textwidth]{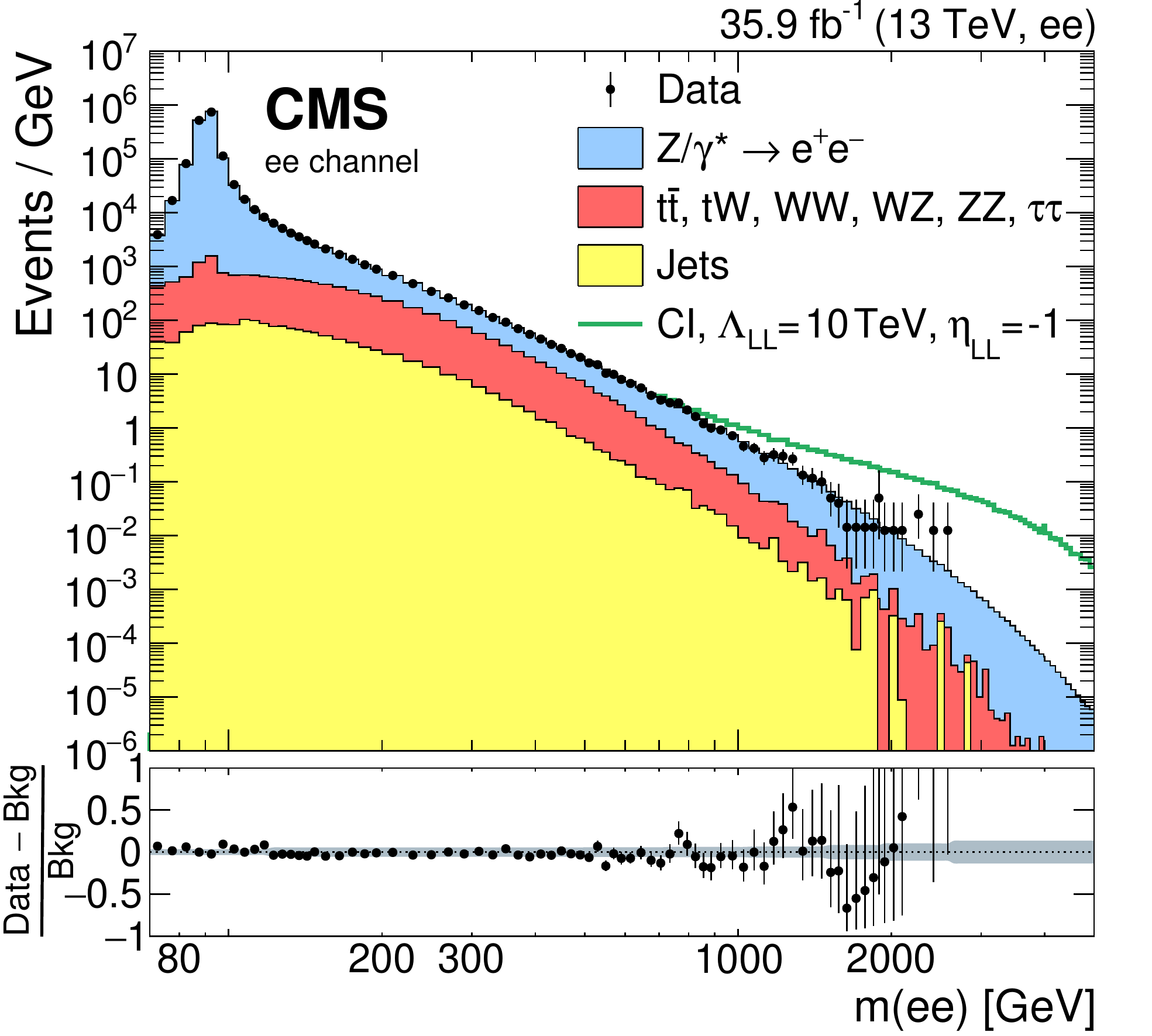}%
  \includegraphics[width=0.5\textwidth]{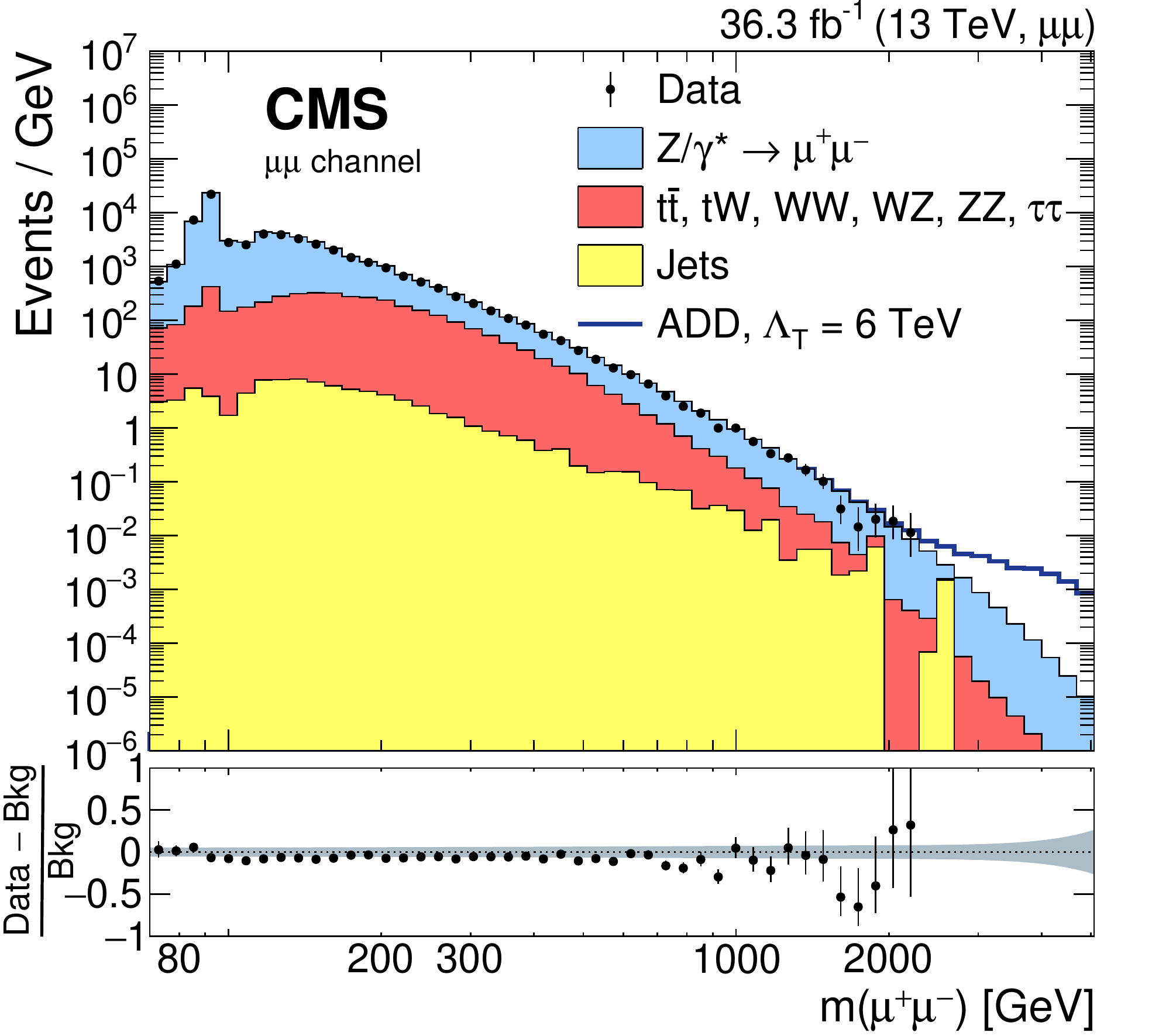}
  \caption{
    Electron (left) and muon (right) pair invariant mass spectra for the combined barrel-barrel and barrel-endcap event categories.
    Example model predictions are given for CI (left) and ADD (right).
    The lower panel shows the relative difference between the data and predicted background.
    The gray band gives the fractional uncertainty (statistical and systematic) in the prediction.
  }
  \label{fig:mass-spectra}
\end{figure}

The resulting dilepton invariant mass spectra for both the electron and muon channels are shown in Fig.~\ref{fig:mass-spectra}, inclusive of the BB and BE event categories.
The simulated events are weighted by the cross section correction factors discussed in Section~\ref{sec:sigbkgdestimation}.
The overall simulated mass distribution is then scaled to fit the observed data yield around the {\cPZ} boson peak ($60<\mll<120\GeV$).

Results from this analysis show no significant deviation from the SM in the dilepton invariant mass spectra for either the electron or muon channel.
Exclusion limits are set on the signal cross section, which are translated into limits on the respective parameters of interest for each model.
These limits are calculated using Bayesian inference, utilizing the framework developed for statistically combining Higgs boson searches~\cite{CMS-NOTE-2011-005}, which is based on the \textsc{RooStats} package~\cite{RooStats}.
All uncertainties are modeled with log-normal probability density functions, while a uniform prior is used for the signal cross section.

\begin{figure}[h]
  \centering
  \includegraphics[width=0.5\textwidth]{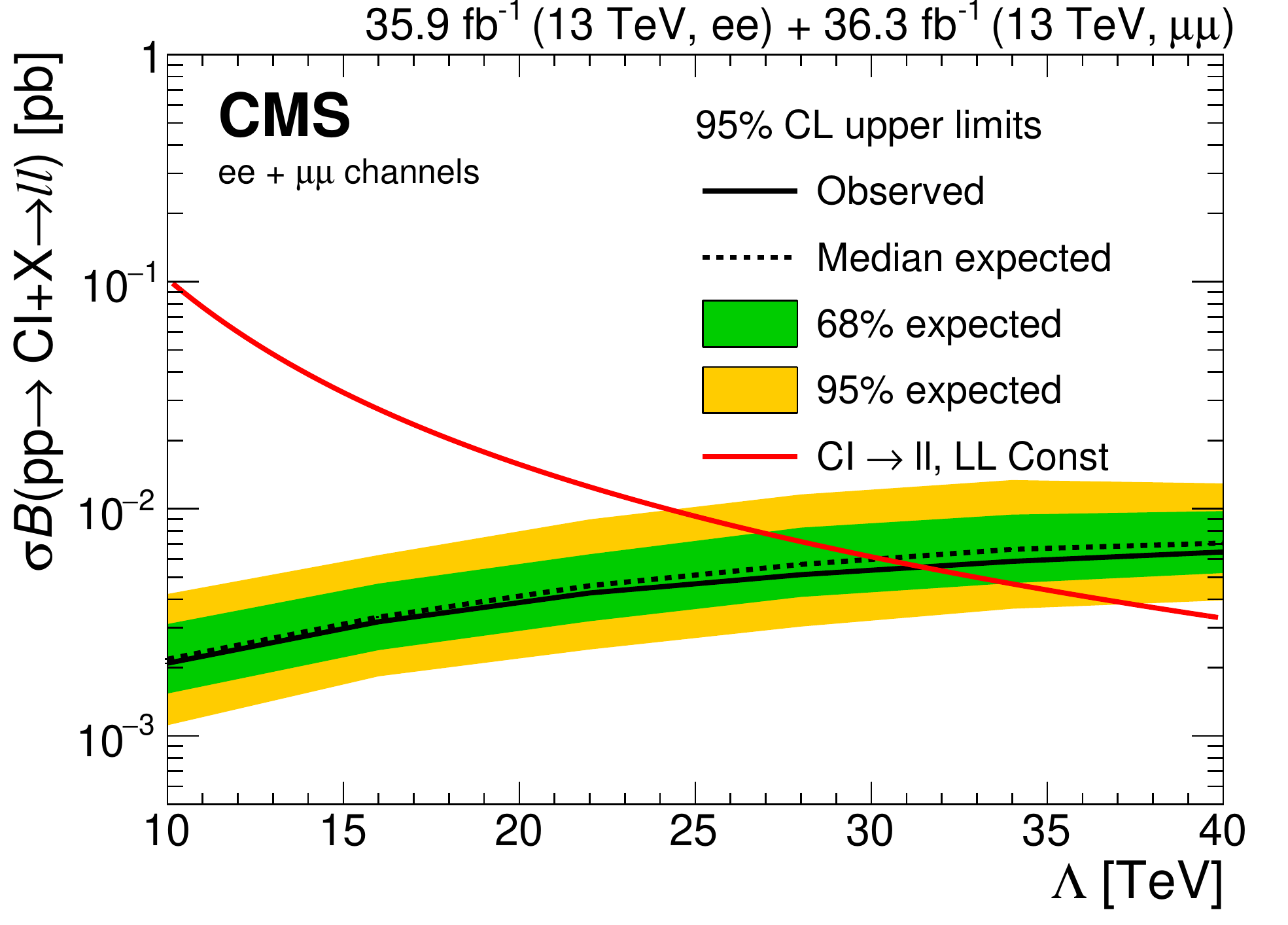}%
  \includegraphics[width=0.5\textwidth]{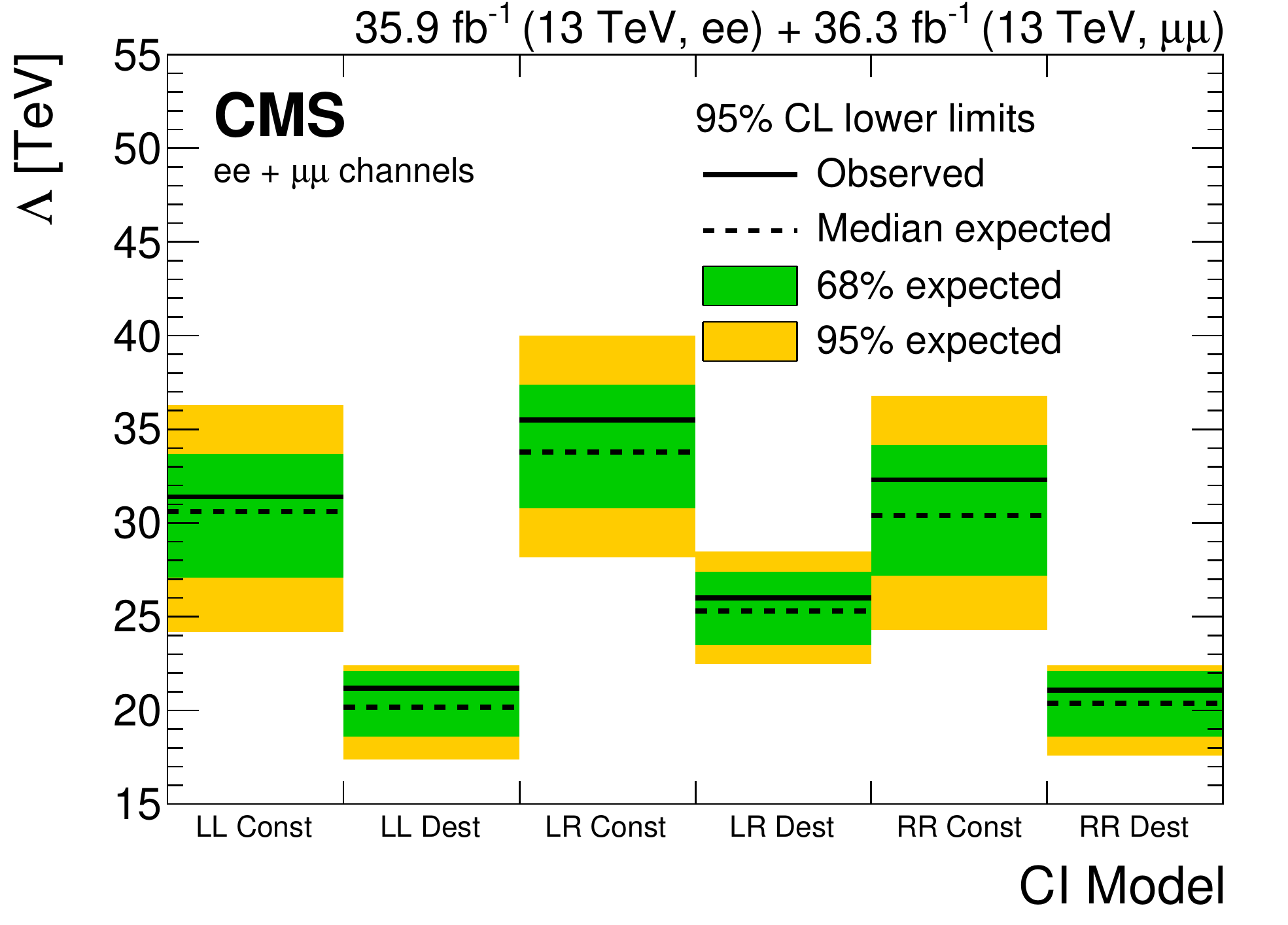}
  \caption{
    Combined dilepton 95\% \CL exclusion limits on the cross section for the left-left constructive CI model (left), and on the CI scale ($\Lambda$) for the six different CI models considered (right).
    The red curve in the left plot shows the theoretical cross section as a function of $\Lambda$.
    The limits are obtained for $\mll>400\,(2200)\GeV$ in the case of constructive (destructive) interference.
  }
  \label{fig:ci-limits-comb}
\end{figure}

For the CI models, two different approaches are used, depending on the signal model.
A single-bin counting experiment with a lower mass threshold of 2.2\TeV, optimized for the best expected limit, is performed for the destructive interference scenarios to remove masses where the signal contribution is negative because of interference with the DY process.
In the case of constructive interference, an alternative approach is used.
The invariant mass spectrum is split into multiple exclusive bins, with lower bin edges of 400, 500, 700, 1100, 1900, and 3500\GeV.
The last bin has an upper edge of 5000\GeV and all bins are combined in the limit calculation.
Systematic uncertainties are treated as fully correlated among the bins.
Expected and observed lower limits on $\Lambda$ are determined from the intersection of the curves for the predicted cross section and the expected and observed upper limits on the CI cross section as a function of $\Lambda$.
This is illustrated in Fig.~\ref{fig:ci-limits-comb} for the left-left constructive model, where the electron and muon channels are combined.

\begin{figure}[h]
  \centering
  \includegraphics[width=0.5\textwidth]{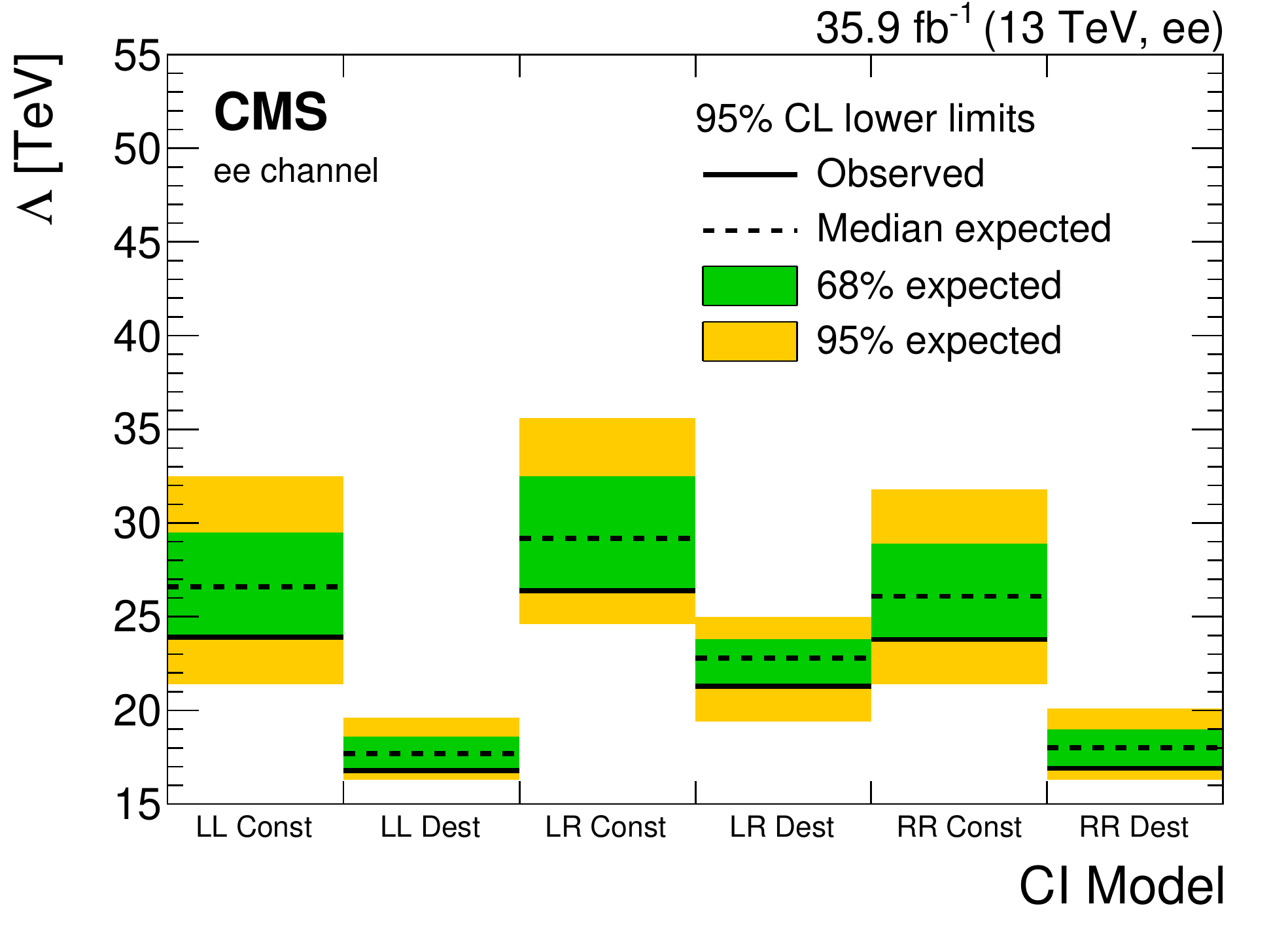}%
  \includegraphics[width=0.5\textwidth]{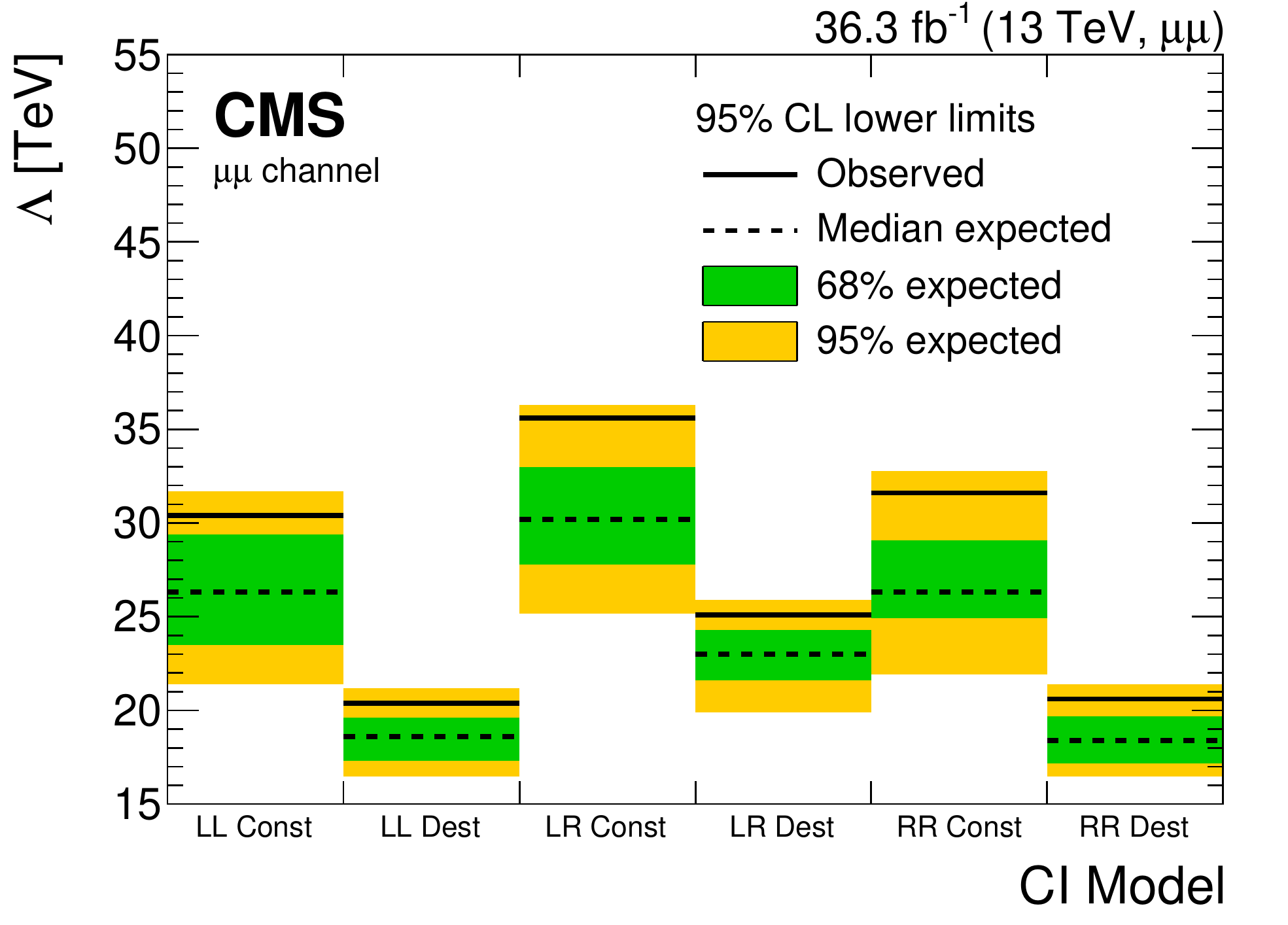}
  \caption{
    Dilepton exclusion limits at 95\% \CL on the CI scale ($\Lambda$) for the six CI models considered for the electron (left) and muon (right) channels.
    The limits are obtained for $\mll>400\,(2200)\GeV$ in the case of constructive (destructive) interference.
  }
  \label{fig:ci-limits}
\end{figure}

The 95\% confidence level (\CL) exclusion limits on the CI model parameter $\Lambda$ are shown in Fig.~\ref{fig:ci-limits} for the six helicity and interference models described in the introduction.
The limits are more stringent for models with constructive interference than those with destructive interference.
The expected limits are comparable for the electron and muon channels, which are shown separately.
The observed limits are more stringent for the muon channel than for the electron channel, but are consistent within statistical fluctuation.
Assuming a universal contact interaction for electrons and muons, exclusion limits can be determined for the combined data sets.
These limits, shown in Fig.~\ref{fig:ci-limits-comb}, range from $\lambdaCIADD{LL}>20\TeV$ for destructive interference to $\lambdaCIADD{RR}>32\TeV$ for constructive interference.

\begin{figure}[h]
  \centering
  \includegraphics[width=0.5\textwidth]{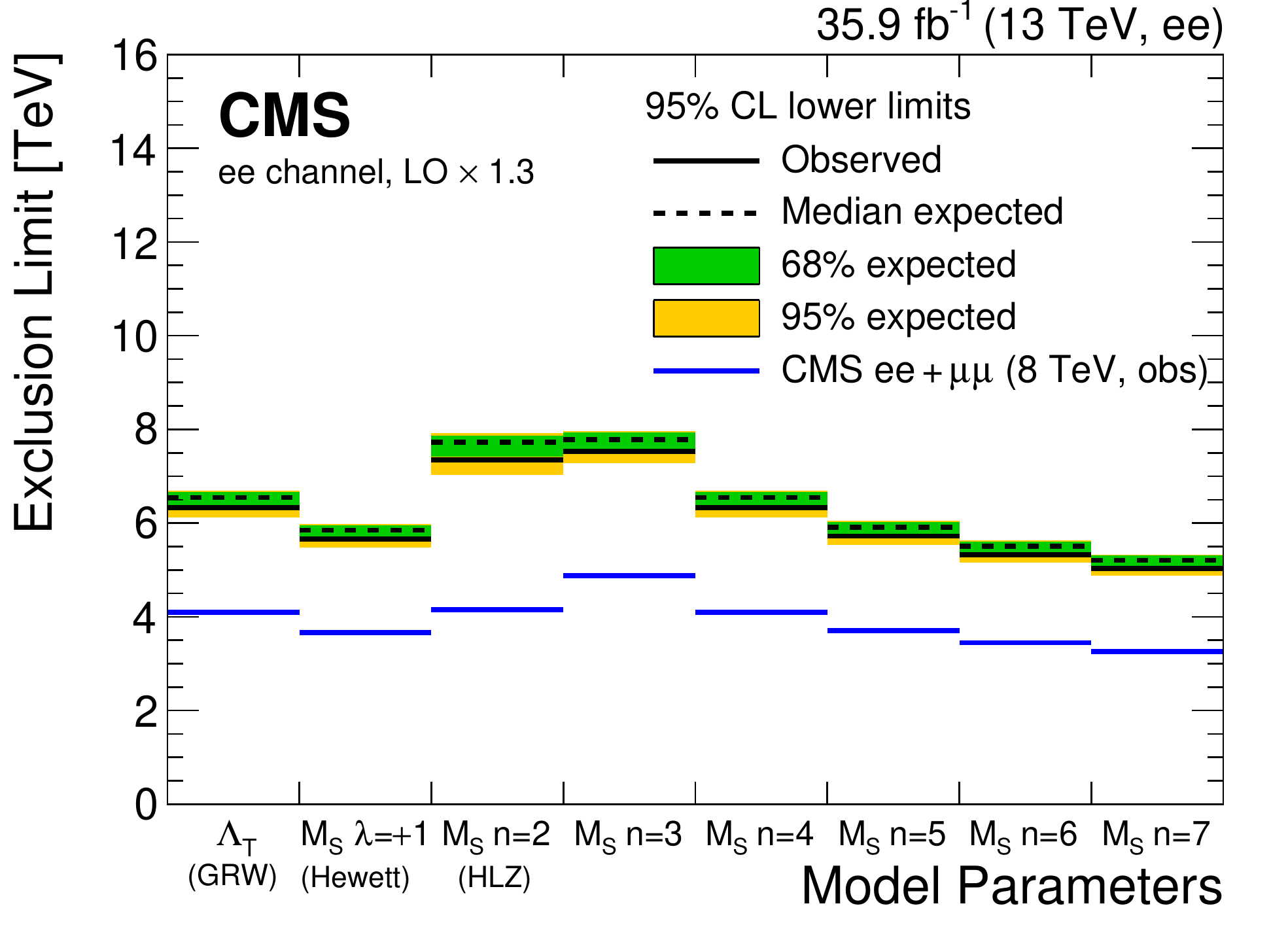}%
  \includegraphics[width=0.5\textwidth]{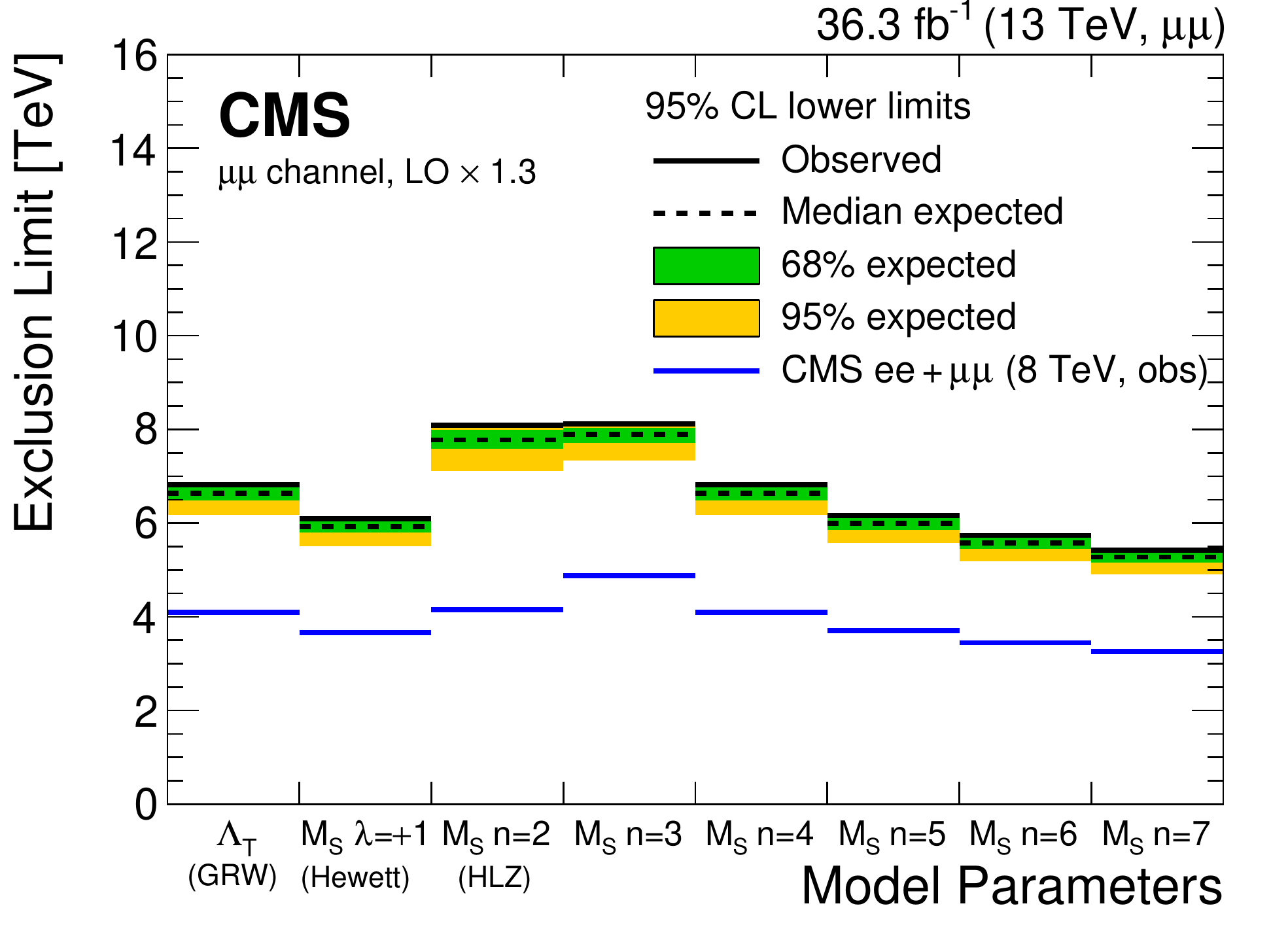}
  \caption{
    Exclusion limits at 95\% \CL on the UV cutoff for the electron (left) and muon (right) channels with $\mll>1.8\TeV$ in the GRW, Hewett, and HLZ conventions for the ADD model.
    Signal model cross sections are calculated up to leading order and a correction factor of 1.3 is applied.
    The results are compared to the previous combined result from CMS~\cite{Khachatryan:2014fba}.
  }
  \label{fig:led-limit-channels}
\end{figure}

\begin{figure}[h]
  \centering
  \includegraphics[width=0.5\textwidth]{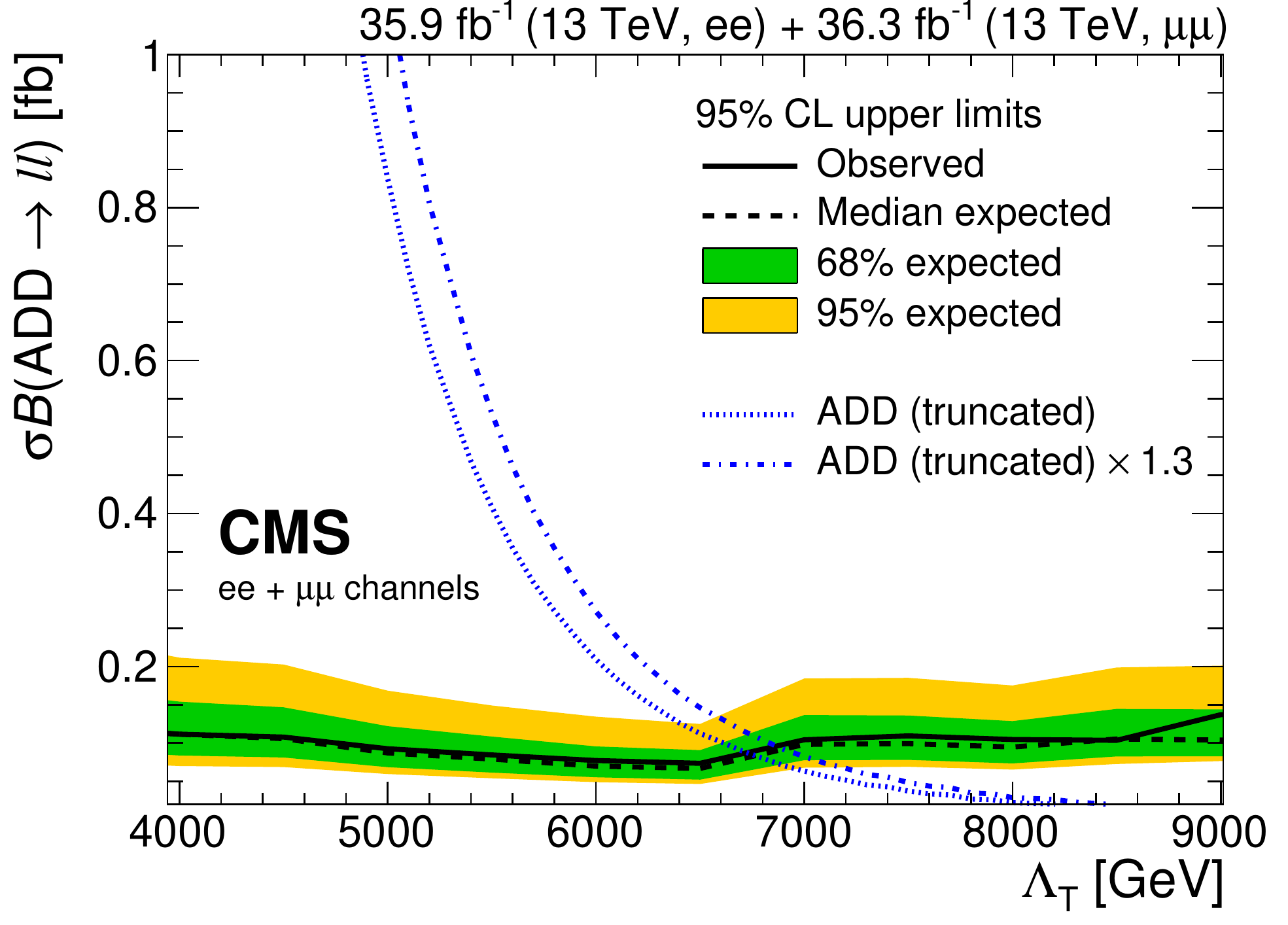}%
  \includegraphics[width=0.5\textwidth]{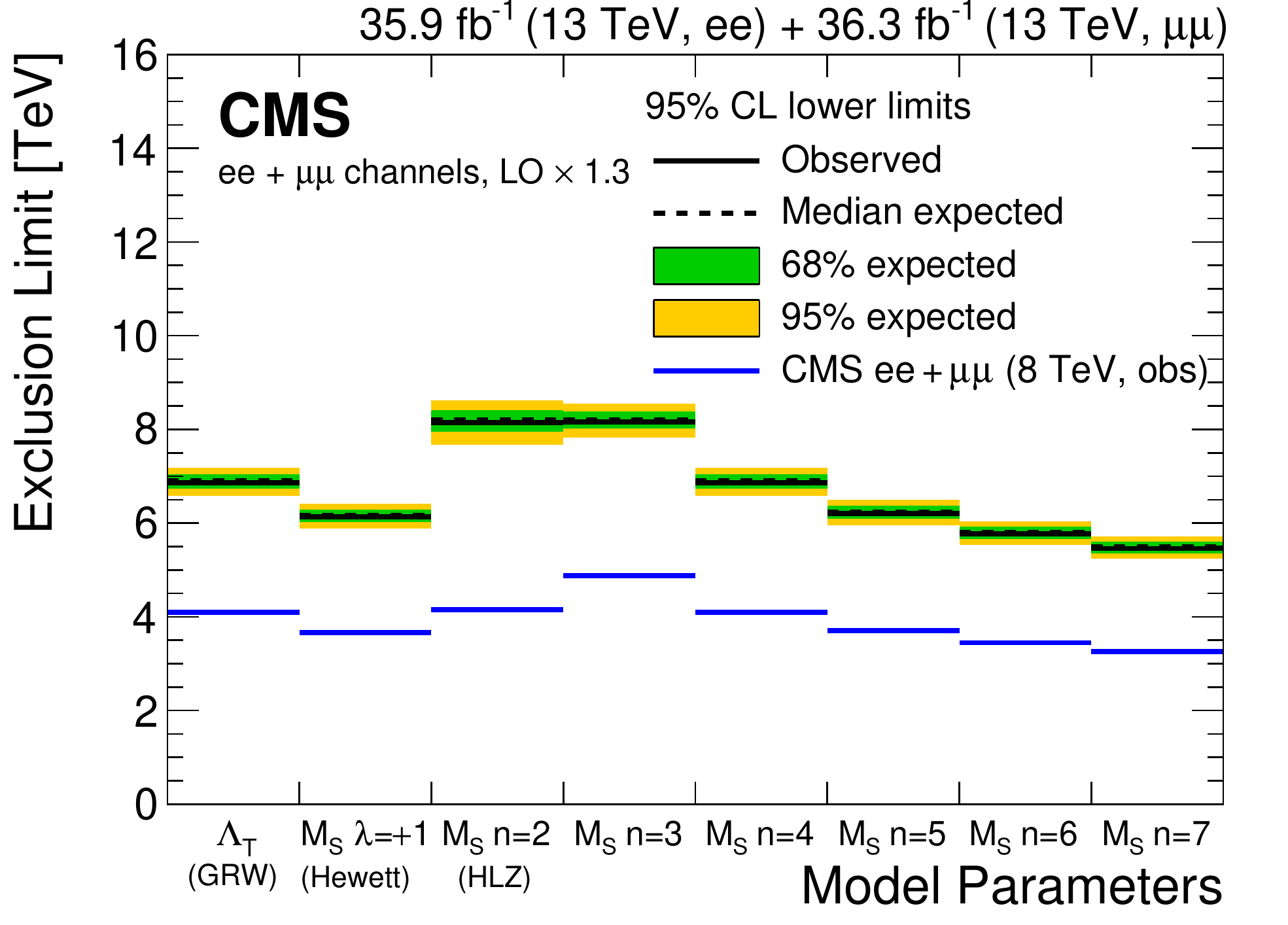}
  \caption{
    Combined dilepton 95\% \CL exclusion limit on the cross section in the GRW convention (left) and on the UV cutoff for all parameter conventions (right) with $\mll>1.8\TeV$ for the ADD model.
    The curves labeled ADD in the left plot show the theoretical signal cross section calculated by \PYTHIA, as a function of the cutoff parameter \lambdaCIADD{T}, and signal contributions with $\mll>\lambdaCIADD{T}$ are set to 0.
    Signal model cross sections are calculated up to leading order and, where indicated by the appropriate label, a correction factor of 1.3 is applied.
    The results are compared to previous ones from CMS~\cite{Khachatryan:2014fba}.
  }
  \label{fig:led-limit-combined}
\end{figure}

For the ADD model, the most sensitive part of the invariant mass spectrum, $\mll>1.8\TeV$, is subdivided into 400\GeV wide search regions, with the final region covering the mass range between 3\TeV and \lambdaCIADD{T}, beyond which all signal contributions are set to 0.
Differentiating between the BB and BE pseudorapidity categories enhances the sensitivity as the signal is expected to be more central than the SM backgrounds.
The most frequently studied parameter conventions, \ie, GRW, Hewett, and HLZ, have been considered.
Figure~\ref{fig:led-limit-channels} shows the 95\% \CL exclusion limits for the respective UV cutoff parameters in both the electron and muon channels.
The combined 95\% \CL exclusion limit on the cross section in the GRW model is shown in Fig.~\ref{fig:led-limit-combined}, alongside the corresponding exclusion limits on the UV cutoff parameters.
The lower limit on \lambdaCIADD{T} at 95\% confidence level is 6.9\TeV, which excludes a string scale \massADD{S} below 6.1\TeV in the Hewett parameter convention.
In the HLZ convention, this translates to lower limits on \massADD{S} of 5.5 to 8.2\TeV, depending on the number of extra dimensions.

\begin{figure}[h]
  \centering
  \includegraphics[width=0.5\textwidth]{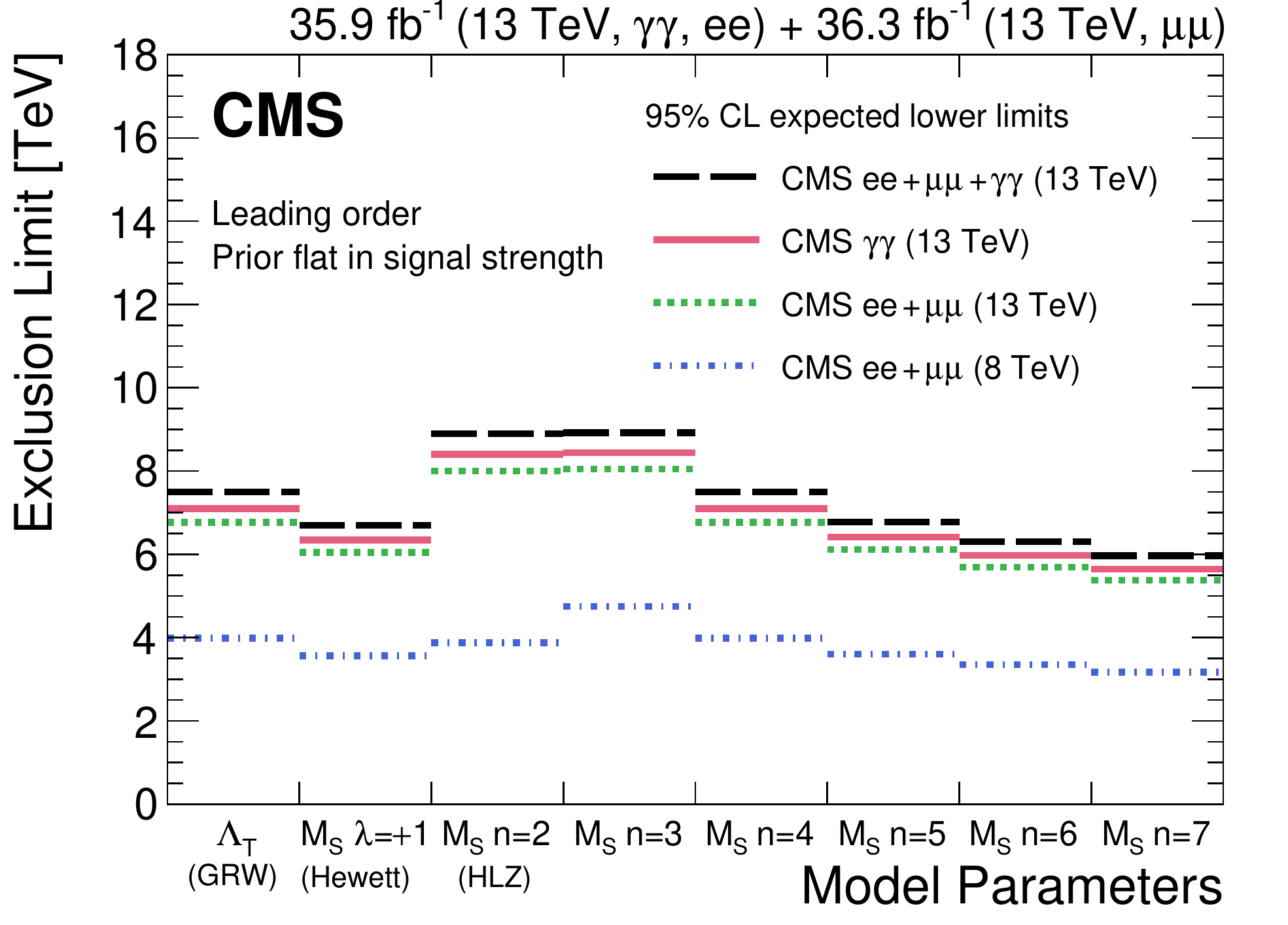}%
  \includegraphics[width=0.5\textwidth]{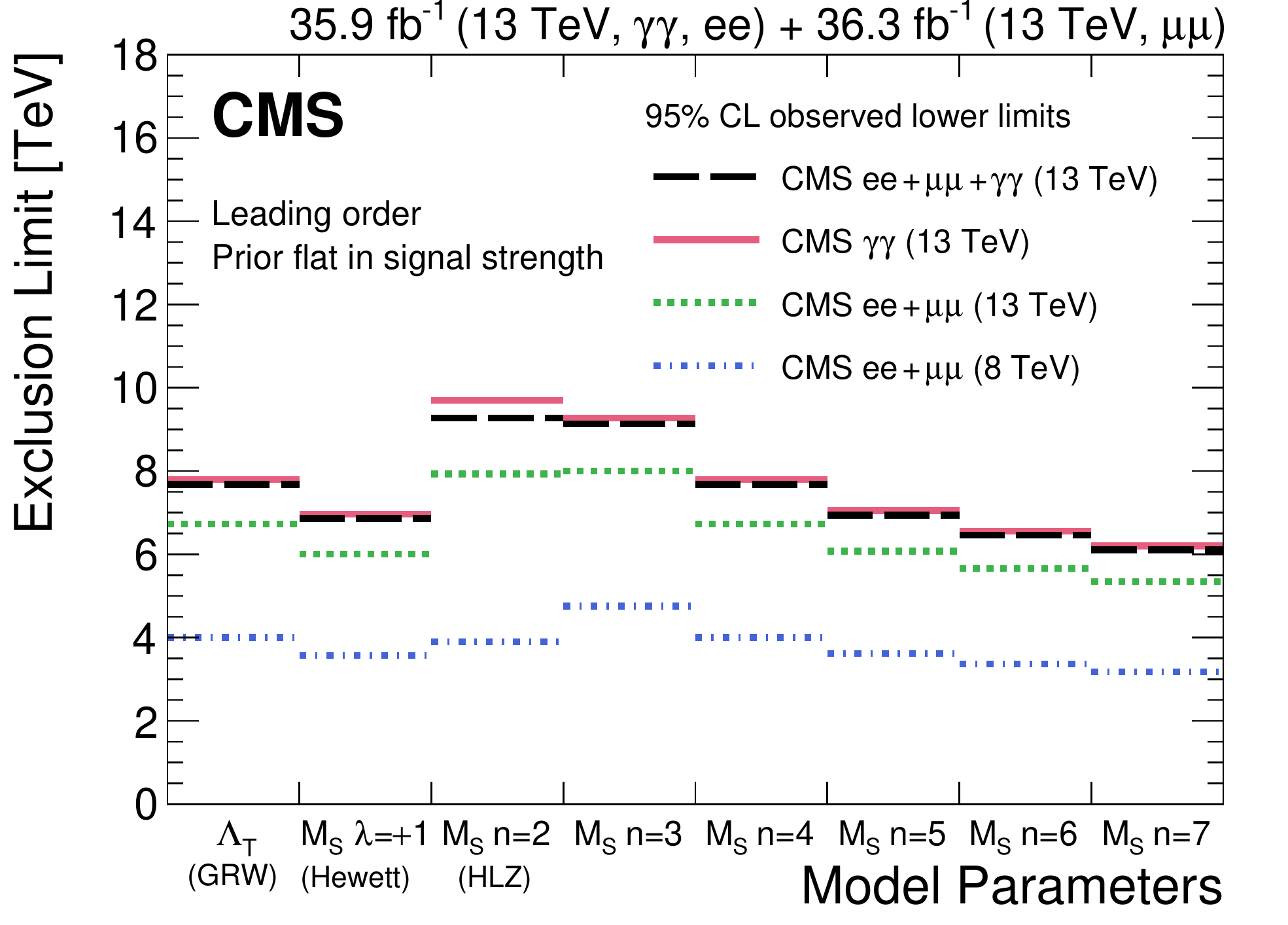}
  \caption{
    Individual and combined dilepton (this analysis) and diphoton~\cite{Sirunyan:2018wnk} 95\% \CL expected (left) and observed (right) exclusion limits as a summary of all parameter conventions for the ADD model.
    Signal model cross sections are calculated up to leading order.
    The dilepton limits from the $\sqrt{s}=8\TeV$ measurement~\cite{Khachatryan:2014fba} are also shown.
  }
  \label{fig:led-limit-dilepton-diphoton}
\end{figure}

Utilizing the recent measurement of diphoton production~\cite{Sirunyan:2018wnk}, the overall sensitivity of the statistical analysis is further improved.
Combining the data of the individual electron, muon, and photon channels, 95\% \CL exclusion limits are calculated using the \textsc{Theta} limit-setting framework~\cite{theta}.
As the scales of the interactions corresponding to the considered search regions, $\mgg>500\GeV$ and $\mll>1.8\TeV$, differ substantially, the uncertainties are taken to be uncorrelated between the diphoton and dilepton analyses.
To ensure a consistent interpretation of the exclusion limits in the combination of all three channels, no higher-order correction factor is assumed.
Figure~\ref{fig:led-limit-dilepton-diphoton} shows the individual and combined limits, and the limits from the $\sqrt{s}=8\TeV$ dilepton measurement~\cite{Khachatryan:2014fba} are also shown.
The highest sensitivity is given by the combination of all three channels as exhibited by the expected limits.
However, an underfluctuation measured in the photon channel still results in the best observed limits.
A summary of the exclusion limits on the respective UV cutoff parameters is given in Table~\ref{tab:led-limit-summary}.
The lower limit on \lambdaCIADD{T} increases to 7.7\TeV, while the limits on \massADD{S} increase to 6.9\TeV in the Hewett convention and 6.1 to 9.3\TeV in the HLZ convention.

\begin{table}[htb]
  \centering
  \topcaption{
    Exclusion limits at 95\% \CL for the electron and muon channels, their combination, and the combination with the diphoton~\cite{Sirunyan:2018wnk} analysis, in multiple parameter conventions of the ADD model.
    Signal model cross sections are calculated up to leading order and, where indicated by the appropriate label, a correction factor of 1.3 is applied.
    For each of the model parameters, the first value is the observed limit followed by the expected limit in parentheses.}
  \cmsTable{
    \begin{tabular}{lcccccccc}
      \hline
      & GRW & Hewett & \multicolumn{6}{c}{HLZ} \\
      \multicolumn{1}{c}{Order} & $\lambdaCIADD{T} [{\TeVns}]$ & $\massADD{S} [{\TeVns}]$ & \multicolumn{6}{c}{$\massADD{S} [{\TeVns}]$} \\
      &  & $\lambda = +1$ & $n = 2$ & $n = 3$ & $n = 4$ & $n = 5$ & $n = 6$ & $n = 7$ \\
      \hline
      \multicolumn{9}{c}{{\Pe\Pe} for $\mee>1.8\TeV$} \\
      LO              & 6.1 (6.4) & 5.5 (5.7) & 7.0 (7.5) & 7.3 (7.6) & 6.1 (6.4) & 5.5 (5.8) & 5.1 (5.4) & 4.9 (5.1) \\
      LO $\times 1.3$ & 6.3 (6.5) & 5.7 (5.8) & 7.3 (7.7) & 7.5 (7.8) & 6.3 (6.5) & 5.7 (5.9) & 5.3 (5.5) & 5.0 (5.2) \\[\cmsTabSkip]
      \multicolumn{9}{c}{{\Pgm\Pgm} for $\muu>1.8\TeV$} \\
      LO              & 6.7 (6.5) & 6.0 (5.8) & 7.9 (7.6) & 7.9 (7.7) & 6.7 (6.5) & 6.0 (5.9) & 5.6 (5.5) & 5.3 (5.2) \\
      LO $\times 1.3$ & 6.8 (6.6) & 6.1 (5.9) & 8.1 (7.8) & 8.1 (7.9) & 6.8 (6.6) & 6.2 (6.0) & 5.7 (5.6) & 5.4 (5.3) \\[\cmsTabSkip]
      \multicolumn{9}{c}{Combined {\Pe\Pe} and {\Pgm\Pgm} for $\mll>1.8\TeV$} \\
      LO              & 6.7 (6.8) & 6.0 (6.0) & 7.9 (8.0) & 8.0 (8.0) & 6.7 (6.8) & 6.1 (6.1) & 5.7 (5.7) & 5.4 (5.4) \\
      LO $\times 1.3$ & 6.9 (6.9) & 6.1 (6.2) & 8.2 (8.2) & 8.2 (8.2) & 6.9 (6.9) & 6.2 (6.2) & 5.8 (5.8) & 5.5 (5.5) \\[\cmsTabSkip]
      \multicolumn{9}{c}{Combined {\Pe\Pe}, {\Pgm\Pgm}, and {\Pgg\Pgg} for $\mll>1.8\TeV$ and $\mgg>500\GeV$} \\
      LO              & 7.7 (7.5) & 6.9 (6.7) & 9.3 (8.9) & 9.1 (8.9) & 7.7 (7.5) & 6.9 (6.8) & 6.5 (6.3) & 6.1 (6.0) \\
      \hline
    \end{tabular}
  }
  \label{tab:led-limit-summary}
\end{table}

\section{Summary}
\label{sec:summary}
A search for nonresonant excesses in the invariant mass spectra of electron and muon pairs has been presented.
The data set recorded with the CMS detector during 2016 is analyzed, corresponding to an integrated luminosity of {\eeLumi} ({\uuLumi})\fbinv for the electron (muon) channel.
No significant deviations from standard model expectations are observed.

A contact interaction (CI) model, taking into account both constructive and destructive interference scenarios, has been used for interpreting the experimental measurements.
The 95\% confidence level exclusion limits on the compositeness scale range from $\lambdaCIADD{LL}>20\TeV$ for the destructive case to $\lambdaCIADD{RR}>32\TeV$ for the constructive one, for the left-left and the right-right helicity currents, respectively.

For the Arkani-Hamed--Dimopoulos--Dvali (ADD) model of large extra dimensions, values of the ultraviolet cutoff parameter \lambdaCIADD{T} (in the Giudice--Rattazzi--Wells, GRW, convention) below 6.9\TeV have been excluded at the 95\% confidence level.
This corresponds to an exclusion on the string scale \massADD{S} below 6.1\TeV in the Hewett convention; in the Han--Lykken--Zhang (HLZ) convention, lower limits are set on \massADD{S} that range from 5.5 to 8.2\TeV, depending on the number of extra dimensions.
When combined with the results from the latest CMS diphoton analysis~\cite{Sirunyan:2018wnk}, these limits improve to 7.7\TeV (GRW), 6.9\TeV (Hewett), and the range 6.1 to 9.3\TeV (HLZ), respectively.

The results presented here for the CI and ADD models improve on previous CMS results at $\sqrt{s}=8\TeV$ in the dilepton final state~\cite{Khachatryan:2014fba}.
The CI limits on $\Lambda$ are compatible with the dilepton results reported by the ATLAS Collaboration~\cite{Aad:2014wca,Aaboud:2017buh}.
However, an exact comparison is not possible because the ATLAS limits are based on priors for $\Lambda$, whereas the limits reported here are based on a prior that is flat in cross section.
For the ADD model, the results reported here are the first measurements at $\sqrt{s}=13\TeV$ in the dilepton final state.
The combination with the CMS diphoton analysis yields the most sensitive results in nonhadronic final states to date.

\begin{acknowledgments}
  We congratulate our colleagues in the CERN accelerator departments for the excellent performance of the LHC and thank the technical and administrative staffs at CERN and at other CMS institutes for their contributions to the success of the CMS effort.
  In addition, we gratefully acknowledge the computing centers and personnel of the Worldwide LHC Computing Grid for delivering so effectively the computing infrastructure essential to our analyses.
  Finally, we acknowledge the enduring support for the construction and operation of the LHC and the CMS detector provided by the following funding agencies: BMBWF and FWF (Austria); FNRS and FWO (Belgium); CNPq, CAPES, FAPERJ, FAPERGS, and FAPESP (Brazil); MES (Bulgaria); CERN; CAS, MoST, and NSFC (China); COLCIENCIAS (Colombia); MSES and CSF (Croatia); RPF (Cyprus); SENESCYT (Ecuador); MoER, ERC IUT, and ERDF (Estonia); Academy of Finland, MEC, and HIP (Finland); CEA and CNRS/IN2P3 (France); BMBF, DFG, and HGF (Germany); GSRT (Greece); NKFIA (Hungary); DAE and DST (India); IPM (Iran); SFI (Ireland); INFN (Italy); MSIP and NRF (Republic of Korea); MES (Latvia); LAS (Lithuania); MOE and UM (Malaysia); BUAP, CINVESTAV, CONACYT, LNS, SEP, and UASLP-FAI (Mexico); MOS (Montenegro); MBIE (New Zealand); PAEC (Pakistan); MSHE and NSC (Poland); FCT (Portugal); JINR (Dubna); MON, RosAtom, RAS, RFBR, and NRC KI (Russia); MESTD (Serbia); SEIDI, CPAN, PCTI, and FEDER (Spain); MOSTR (Sri Lanka); Swiss Funding Agencies (Switzerland); MST (Taipei); ThEPCenter, IPST, STAR, and NSTDA (Thailand); TUBITAK and TAEK (Turkey); NASU and SFFR (Ukraine); STFC (United Kingdom); DOE and NSF (USA).

  \hyphenation{Rachada-pisek} Individuals have received support from the Marie-Curie programme and the European Research Council and Horizon 2020 Grant, contract No.\ 675440 (European Union); the Leventis Foundation; the A.P.\ Sloan Foundation; the Alexander von Humboldt Foundation; the Belgian Federal Science Policy Office; the Fonds pour la Formation \`a la Recherche dans l'Industrie et dans l'Agriculture (FRIA-Belgium); the Agentschap voor Innovatie door Wetenschap en Technologie (IWT-Belgium); the F.R.S.-FNRS and FWO (Belgium) under the ``Excellence of Science -- EOS" -- be.h project n.\ 30820817; the Ministry of Education, Youth and Sports (MEYS) of the Czech Republic; the Lend\"ulet (``Momentum") Programme and the J\'anos Bolyai Research Scholarship of the Hungarian Academy of Sciences, the New National Excellence Program \'UNKP, the NKFIA research grants 123842, 123959, 124845, 124850, and 125105 (Hungary); the Council of Science and Industrial Research, India; the HOMING PLUS programme of the Foundation for Polish Science, cofinanced from European Union, Regional Development Fund, the Mobility Plus programme of the Ministry of Science and Higher Education, the National Science Center (Poland), contracts Harmonia 2014/14/M/ST2/00428, Opus 2014/13/B/ST2/02543, 2014/15/B/ST2/03998, and 2015/19/B/ST2/02861, Sonata-bis 2012/07/E/ST2/01406; the National Priorities Research Program by Qatar National Research Fund; the Programa Estatal de Fomento de la Investigaci{\'o}n Cient{\'i}fica y T{\'e}cnica de Excelencia Mar\'{\i}a de Maeztu, grant MDM-2015-0509 and the Programa Severo Ochoa del Principado de Asturias; the Thalis and Aristeia programmes cofinanced by EU-ESF and the Greek NSRF; the Rachadapisek Sompot Fund for Postdoctoral Fellowship, Chulalongkorn University and the Chulalongkorn Academic into Its 2nd Century Project Advancement Project (Thailand); the Welch Foundation, contract C-1845; and the Weston Havens Foundation (USA).
\end{acknowledgments}

\bibliography{auto_generated}

\providecommand{\href}[2]{#2}\begingroup\raggedright\begin{thebibliography}{10}%
\makeatletter
\providecommand{\hrefCMSnoop }[0]{\@secondoftwo}%
\makeatother
\providecommand{\doi}{\texttt{doi:}\begingroup \urlstyle{tt}\Url}

\bibitem{Eichten:1984eu}
\hrefCMSnoop {}{E.~Eichten, I.~Hinchliffe, K.~Lane, and C.~Quigg,
  ``Supercollider physics'',} \textit{ Rev. Mod. Phys.} \textbf{ 56} (1984)
  579,
  \href{http://dx.doi.org/10.1103/RevModPhys.56.579}{\doi{10.1103/RevModPhys.56.579}}.
[Erratum: \DOI{10.1103/RevModPhys.58.1065}].
%%CITATION = RMPHA,56,579;%%.

\bibitem{arkani98:hlz}
\hrefCMSnoop {}{N.~Arkani-Hamed, S.~Dimopoulos, and G.~Dvali, ``The hierarchy
  problem and new dimensions at a millimeter'',} \textit{ Phys. Lett. B}
  \textbf{ 429} (1998) 263,
  \href{http://dx.doi.org/10.1016/S0370-2693(98)00466-3}{\doi{10.1016/S0370-2693(98)00466-3}},
  \href{http://www.arXiv.org/abs/hep-ph/9803315}{\texttt{arXiv:hep-ph/9803315}}.

\bibitem{Eichten:1983hw}
\hrefCMSnoop {}{E.~Eichten, K.~Lane, and M.~Peskin, ``New tests for quark and
  lepton substructure'',} \textit{ Phys. Rev. Lett.} \textbf{ 50} (1983) 811,
\href{http://dx.doi.org/10.1103/PhysRevLett.50.811}{\doi{10.1103/PhysRevLett.50.811}}.
%%CITATION = PRLTA,50,811;%%.

\bibitem{PDG2018}
\hrefCMSnoop {}{{Particle Data Group}, M.~Tanabashi {et~al.}, ``Review of
  particle physics'',} \textit{ Phys. Rev. D} \textbf{ 98} (2018) 030001,
  \href{http://dx.doi.org/10.1103/PhysRevD.98.030001}{\doi{10.1103/PhysRevD.98.030001}}.

\bibitem{GRW99:extradim}
\hrefCMSnoop {}{G.~F. Giudice, R.~Rattazzi, and J.~D. Wells, ``Quantum gravity
  and extra dimensions at high-energy colliders'',} \textit{ Nucl. Phys. B}
  \textbf{ 544} (1999) 3,
  \href{http://dx.doi.org/10.1016/S0550-3213(99)00044-9}{\doi{10.1016/S0550-3213(99)00044-9}},
  \href{http://www.arXiv.org/abs/hep-ph/9811291}{\texttt{arXiv:hep-ph/9811291}}.

\bibitem{Hewett:1998sn}
\hrefCMSnoop {}{J.~L. Hewett, ``Indirect collider signals for extra
  dimensions'',} \textit{ Phys. Rev. Lett.} \textbf{ 82} (1999) 4765,
  \href{http://dx.doi.org/10.1103/PhysRevLett.82.4765}{\doi{10.1103/PhysRevLett.82.4765}},
\href{http://www.arXiv.org/abs/hep-ph/9811356}{\texttt{arXiv:hep-ph/9811356}}.
%%CITATION = HEP-PH/9811356;%%.

\bibitem{han99:hlz}
\hrefCMSnoop {}{T.~Han, J.~D. Lykken, and R.~Zhang, ``{K}aluza--{K}lein states
  from large extra dimensions'',} \textit{ Phys. Rev. D} \textbf{ 59} (1999)
  105006,
  \href{http://dx.doi.org/10.1103/PhysRevD.59.105006}{\doi{10.1103/PhysRevD.59.105006}},
  \href{http://www.arXiv.org/abs/hep-ph/9811350}{\texttt{arXiv:hep-ph/9811350}}.

\bibitem{Khachatryan:2014fba}
\hrefCMSnoop {}{{CMS Collaboration}, ``Search for physics beyond the standard
  model in dilepton mass spectra in proton-proton collisions at
  $\sqrt{s}=8$~{TeV}'',} \textit{ JHEP} \textbf{ 04} (2015) 025,
  \href{http://dx.doi.org/10.1007/JHEP04(2015)025}{\doi{10.1007/JHEP04(2015)025}},
\href{http://www.arXiv.org/abs/1412.6302}{\texttt{arXiv:1412.6302}}.
%%CITATION = ARXIV:1412.6302;%%.

\bibitem{Sirunyan:2018exx}
\hrefCMSnoop {}{{CMS Collaboration}, ``Search for high-mass resonances in
  dilepton final states in proton-proton collisions at $\sqrt{s}=13$~{TeV}'',}
  \textit{ JHEP} \textbf{ 06} (2018) 120,
  \href{http://dx.doi.org/10.1007/JHEP06(2018)120}{\doi{10.1007/JHEP06(2018)120}},
\href{http://www.arXiv.org/abs/1803.06292}{\texttt{arXiv:1803.06292}}.
%%CITATION = ARXIV:1803.06292;%%.

\bibitem{Sirunyan:2018wnk}
\hrefCMSnoop {}{{CMS Collaboration}, ``Search for physics beyond the standard
  model in high-mass diphoton events from proton-proton collisions at
  $\sqrt{s}=13$~{TeV}'',} \textit{ Phys. Rev. D} \textbf{ 98} (2018) 092001,
  \href{http://dx.doi.org/10.1103/PhysRevD.98.092001}{\doi{10.1103/PhysRevD.98.092001}},
\href{http://www.arXiv.org/abs/1809.00327}{\texttt{arXiv:1809.00327}}.
%%CITATION = ARXIV:1809.00327;%%.

\bibitem{Sirunyan:2018wcm}
\hrefCMSnoop {}{{CMS Collaboration}, ``Search for new physics in dijet angular
  distributions using proton-proton collisions at $\sqrt{s}=13$~{TeV} and
  constraints on dark matter and other models'',} \textit{ Eur. Phys. J. C}
  \textbf{ 78} (2018) 789,
  \href{http://dx.doi.org/10.1140/epjc/s10052-018-6242-x}{\doi{10.1140/epjc/s10052-018-6242-x}},
\href{http://www.arXiv.org/abs/1803.08030}{\texttt{arXiv:1803.08030}}.
%%CITATION = ARXIV:1803.08030;%%.

\bibitem{Aad:2014wca}
\hrefCMSnoop {}{{ATLAS Collaboration}, ``Search for contact interactions and
  large extra dimensions in the dilepton channel using proton-proton collisions
  at $\sqrt{s}=8$~{TeV} with the {ATLAS} detector'',} \textit{ Eur. Phys. J. C}
  \textbf{ 74} (2014) 3134,
  \href{http://dx.doi.org/10.1140/epjc/s10052-014-3134-6}{\doi{10.1140/epjc/s10052-014-3134-6}},
\href{http://www.arXiv.org/abs/1407.2410}{\texttt{arXiv:1407.2410}}.
%%CITATION = ARXIV:1407.2410;%%.

\bibitem{Aaboud:2017buh}
\hrefCMSnoop {}{{ATLAS Collaboration}, ``Search for new high-mass phenomena in
  the dilepton final state using 36 fb$^{-1}$ of proton-proton collision data
  at $\sqrt{s}=13$~{TeV} with the {ATLAS} detector'',} \textit{ JHEP} \textbf{
  10} (2017) 182,
  \href{http://dx.doi.org/10.1007/JHEP10(2017)182}{\doi{10.1007/JHEP10(2017)182}},
\href{http://www.arXiv.org/abs/1707.02424}{\texttt{arXiv:1707.02424}}.
%%CITATION = ARXIV:1707.02424;%%.

\bibitem{CMS-JINST}
\hrefCMSnoop {}{{CMS Collaboration}, ``The {CMS} experiment at the {CERN}
  {LHC}'',} \textit{ JINST} \textbf{ 3} (2008) S08004,
\href{http://dx.doi.org/10.1088/1748-0221/3/08/S08004}{\doi{10.1088/1748-0221/3/08/S08004}}.
%%CITATION = JINST,3,S08004;%%.

\bibitem{Khachatryan:2016bia}
\hrefCMSnoop {}{{CMS Collaboration}, ``The {CMS} trigger system'',} \textit{
  JINST} \textbf{ 12} (2017) P01020,
  \href{http://dx.doi.org/10.1088/1748-0221/12/01/P01020}{\doi{10.1088/1748-0221/12/01/P01020}},
\href{http://www.arXiv.org/abs/1609.02366}{\texttt{arXiv:1609.02366}}.
%%CITATION = ARXIV:1609.02366;%%.

\bibitem{Khachatryan:2016zqb}
\hrefCMSnoop {}{{CMS Collaboration}, ``Search for narrow resonances in dilepton
  mass spectra in proton-proton collisions at $\sqrt{s}=13$~{TeV} and
  combination with 8 {TeV} data'',} \textit{ Phys. Lett. B} \textbf{ 768}
  (2017) 57,
  \href{http://dx.doi.org/10.1016/j.physletb.2017.02.010}{\doi{10.1016/j.physletb.2017.02.010}},
\href{http://www.arXiv.org/abs/1609.05391}{\texttt{arXiv:1609.05391}}.
%%CITATION = ARXIV:1609.05391;%%.

\bibitem{Khachatryan:2015hwa}
\hrefCMSnoop {}{{CMS Collaboration}, ``Performance of electron reconstruction
  and selection with the {CMS} detector in proton-proton collisions at
  $\sqrt{s}=8$~{TeV}'',} \textit{ JINST} \textbf{ 10} (2015) P06005,
  \href{http://dx.doi.org/10.1088/1748-0221/10/06/P06005}{\doi{10.1088/1748-0221/10/06/P06005}},
\href{http://www.arXiv.org/abs/1502.02701}{\texttt{arXiv:1502.02701}}.
%%CITATION = ARXIV:1502.02701;%%.

\bibitem{Sirunyan:2018fpa}
\hrefCMSnoop {}{{CMS Collaboration}, ``Performance of the {CMS} muon detector
  and muon reconstruction with proton-proton collisions at
  $\sqrt{s}=13$~{TeV}'',} \textit{ JINST} \textbf{ 13} (2018) P06015,
  \href{http://dx.doi.org/10.1088/1748-0221/13/06/P06015}{\doi{10.1088/1748-0221/13/06/P06015}},
\href{http://www.arXiv.org/abs/1804.04528}{\texttt{arXiv:1804.04528}}.
%%CITATION = ARXIV:1804.04528;%%.

\bibitem{Nason:2004rx}
\hrefCMSnoop {}{P.~Nason, ``A new method for combining {NLO} {QCD} with shower
  {M}onte {C}arlo algorithms'',} \textit{ JHEP} \textbf{ 11} (2004) 040,
  \href{http://dx.doi.org/10.1088/1126-6708/2004/11/040}{\doi{10.1088/1126-6708/2004/11/040}},
\href{http://www.arXiv.org/abs/hep-ph/0409146}{\texttt{arXiv:hep-ph/0409146}}.
%%CITATION = HEP-PH/0409146;%%.

\bibitem{Frixione:2007vw}
\hrefCMSnoop {}{S.~Frixione, P.~Nason, and C.~Oleari, ``Matching {NLO} {QCD}
  computations with parton shower simulations: the {POWHEG} method'',} \textit{
  JHEP} \textbf{ 11} (2007) 070,
  \href{http://dx.doi.org/10.1088/1126-6708/2007/11/070}{\doi{10.1088/1126-6708/2007/11/070}},
\href{http://www.arXiv.org/abs/0709.2092}{\texttt{arXiv:0709.2092}}.
%%CITATION = 0709.2092;%%.

\bibitem{Alioli:2010xd}
\hrefCMSnoop {}{S.~Alioli, P.~Nason, C.~Oleari, and E.~Re, ``A general
  framework for implementing {NLO} calculations in shower {M}onte {C}arlo
  programs: the {POWHEG} {BOX}'',} \textit{ JHEP} \textbf{ 06} (2010) 043,
  \href{http://dx.doi.org/10.1007/JHEP06(2010)043}{\doi{10.1007/JHEP06(2010)043}},
\href{http://www.arXiv.org/abs/1002.2581}{\texttt{arXiv:1002.2581}}.
%%CITATION = ARXIV:1002.2581;%%.

\bibitem{Alioli:2008gx}
\hrefCMSnoop {}{S.~Alioli, P.~Nason, C.~Oleari, and E.~Re, ``{NLO} vector-boson
  production matched with shower in {POWHEG}'',} \textit{ JHEP} \textbf{ 07}
  (2008) 060,
  \href{http://dx.doi.org/10.1088/1126-6708/2008/07/060}{\doi{10.1088/1126-6708/2008/07/060}},
\href{http://www.arXiv.org/abs/0805.4802}{\texttt{arXiv:0805.4802}}.
%%CITATION = ARXIV:0805.4802;%%.

\bibitem{Frixione:2007nw}
\hrefCMSnoop {}{S.~Frixione, P.~Nason, and G.~Ridolfi, ``A positive-weight
  next-to-leading-order {M}onte {C}arlo for heavy flavour hadroproduction'',}
  \textit{ JHEP} \textbf{ 09} (2007) 126,
  \href{http://dx.doi.org/10.1088/1126-6708/2007/09/126}{\doi{10.1088/1126-6708/2007/09/126}},
\href{http://www.arXiv.org/abs/0707.3088}{\texttt{arXiv:0707.3088}}.
%%CITATION = ARXIV:0707.3088;%%.

\bibitem{Re:2010bp}
\hrefCMSnoop {}{E.~Re, ``Single-top {$\PW\cPqt$}-channel production matched
  with parton showers using the {POWHEG} method'',} \textit{ Eur. Phys. J. C}
  \textbf{ 71} (2011) 1547,
  \href{http://dx.doi.org/10.1140/epjc/s10052-011-1547-z}{\doi{10.1140/epjc/s10052-011-1547-z}},
\href{http://www.arXiv.org/abs/1009.2450}{\texttt{arXiv:1009.2450}}.
%%CITATION = ARXIV:1009.2450;%%.

\bibitem{Ball:2014uwa}
\hrefCMSnoop {}{{NNPDF} Collaboration, ``Parton distributions for the {LHC Run
  II}'',} \textit{ JHEP} \textbf{ 04} (2015) 040,
  \href{http://dx.doi.org/10.1007/JHEP04(2015)040}{\doi{10.1007/JHEP04(2015)040}},
\href{http://www.arXiv.org/abs/1410.8849}{\texttt{arXiv:1410.8849}}.
%%CITATION = ARXIV:1410.8849;%%.

\bibitem{Sjostrand:2014zea}
T.~Sj{\"o}strand\hrefCMSnoop {}{ {et~al.}, ``An introduction to {PYTHIA}
  8.2'',} \textit{ Comput. Phys. Commun.} \textbf{ 191} (2015) 159,
  \href{http://dx.doi.org/10.1016/j.cpc.2015.01.024}{\doi{10.1016/j.cpc.2015.01.024}},
\href{http://www.arXiv.org/abs/1410.3012}{\texttt{arXiv:1410.3012}}.
%%CITATION = ARXIV:1410.3012;%%.

\bibitem{Li:2012wna}
\hrefCMSnoop {}{Y.~Li and F.~Petriello, ``Combining {QCD} and electroweak
  corrections to dilepton production in the framework of the {FEWZ} simulation
  code'',} \textit{ Phys. Rev. D} \textbf{ 86} (2012) 094034,
  \href{http://dx.doi.org/10.1103/PhysRevD.86.094034}{\doi{10.1103/PhysRevD.86.094034}},
  \href{http://www.arXiv.org/abs/1208.5967}{\texttt{arXiv:1208.5967}}.

\bibitem{Botje:2011sn}
M.~Botje\hrefCMSnoop {}{ {et~al.}, ``{The PDF4LHC Working Group Interim
  Recommendations}'',} (2011).
\href{http://www.arXiv.org/abs/1101.0538}{\texttt{arXiv:1101.0538}}.
%%CITATION = ARXIV:1101.0538;%%.

\bibitem{Alekhin:2011sk}
\hrefCMSnoop {}{S.~Alekhin {et~al.}, ``{The PDF4LHC Working Group Interim
  Report}'',} (2011).
\href{http://www.arXiv.org/abs/1101.0536}{\texttt{arXiv:1101.0536}}.
%%CITATION = ARXIV:1101.0536;%%.

\bibitem{Butterworth:2015oua}
\hrefCMSnoop {}{J.~Butterworth {et~al.}, ``{PDF4LHC} recommendations for {LHC
  Run II}'',} \textit{ J. Phys. G} \textbf{ 43} (2016) 023001,
  \href{http://dx.doi.org/10.1088/0954-3899/43/2/023001}{\doi{10.1088/0954-3899/43/2/023001}},
\href{http://www.arXiv.org/abs/1510.03865}{\texttt{arXiv:1510.03865}}.
%%CITATION = ARXIV:1510.03865;%%.

\bibitem{Manohar:2016nzj}
\hrefCMSnoop {}{A.~Manohar, P.~Nason, G.~P. Salam, and G.~Zanderighi, ``How
  bright is the proton? {A} precise determination of the photon parton
  distribution function'',} \textit{ Phys. Rev. Lett.} \textbf{ 117} (2016)
  242002,
  \href{http://dx.doi.org/10.1103/PhysRevLett.117.242002}{\doi{10.1103/PhysRevLett.117.242002}},
\href{http://www.arXiv.org/abs/1607.04266}{\texttt{arXiv:1607.04266}}.
%%CITATION = ARXIV:1607.04266;%%.

\bibitem{Bourilkov:2016qum}
\hrefCMSnoop {}{D.~Bourilkov, ``Photon-induced background for dilepton searches
  and measurements in pp collisions at 13 {TeV}'',} (2016).
\href{http://www.arXiv.org/abs/1606.00523}{\texttt{arXiv:1606.00523}}.
%%CITATION = ARXIV:1606.00523;%%.

\bibitem{Bourilkov:2016oet}
\hrefCMSnoop {}{D.~Bourilkov, ``Exploring the {LHC} landscape with
  dileptons'',} (2016).
\href{http://www.arXiv.org/abs/1609.08994}{\texttt{arXiv:1609.08994}}.
%%CITATION = ARXIV:1609.08994;%%.

\bibitem{Agostinelli:2002hh}
\hrefCMSnoop {}{{{GEANT4}} Collaboration, ``{\GEANTfour}---a simulation
  toolkit'',} \textit{ Nucl. Instrum. Meth. A} \textbf{ 506} (2003) 250,
\href{http://dx.doi.org/10.1016/S0168-9002(03)01368-8}{\doi{10.1016/S0168-9002(03)01368-8}}.
%%CITATION = NUIMA,A506,250;%%.

\bibitem{Czakon:2011xx}
\hrefCMSnoop {}{M.~Czakon and A.~Mitov, ``Top++: A program for the calculation
  of the top-pair cross-section at hadron colliders'',} \textit{ Comput. Phys.
  Commun.} \textbf{ 185} (2014) 2930,
  \href{http://dx.doi.org/10.1016/j.cpc.2014.06.021}{\doi{10.1016/j.cpc.2014.06.021}},
\href{http://www.arXiv.org/abs/1112.5675}{\texttt{arXiv:1112.5675}}.
%%CITATION = ARXIV:1112.5675;%%.

\bibitem{Kidonakis:2010ux}
\hrefCMSnoop {}{N.~Kidonakis, ``Two-loop soft anomalous dimensions for single
  top quark associated production with a {W$^{-}$} or {H$^{-}$}'',} \textit{
  Phys. Rev. D} \textbf{ 82} (2010) 054018,
  \href{http://dx.doi.org/10.1103/PhysRevD.82.054018}{\doi{10.1103/PhysRevD.82.054018}},
\href{http://www.arXiv.org/abs/1005.4451}{\texttt{arXiv:1005.4451}}.
%%CITATION = ARXIV:1005.4451;%%.

\bibitem{Boughezal:2016wmq}
R.~Boughezal\hrefCMSnoop {}{ {et~al.}, ``Color-singlet production at {NNLO} in
  {MCFM}'',} \textit{ Eur. Phys. J. C} \textbf{ 77} (2016) 7,
  \href{http://dx.doi.org/10.1140/epjc/s10052-016-4558-y}{\doi{10.1140/epjc/s10052-016-4558-y}},
\href{http://www.arXiv.org/abs/1605.08011}{\texttt{arXiv:1605.08011}}.
%%CITATION = ARXIV:1605.08011;%%.

\bibitem{Campbell:2015qma}
\hrefCMSnoop {}{J.~M. Campbell, R.~K. Ellis, and W.~T. Giele, ``A
  multi-threaded version of {MCFM}'',} \textit{ Eur. Phys. J. C} \textbf{ 75}
  (2015) 246,
  \href{http://dx.doi.org/10.1140/epjc/s10052-015-3461-2}{\doi{10.1140/epjc/s10052-015-3461-2}},
\href{http://www.arXiv.org/abs/1503.06182}{\texttt{arXiv:1503.06182}}.
%%CITATION = ARXIV:1503.06182;%%.

\bibitem{Campbell:2011bn}
\hrefCMSnoop {}{J.~M. Campbell, R.~K. Ellis, and C.~Williams, ``Vector boson
  pair production at the {LHC}'',} \textit{ JHEP} \textbf{ 07} (2011) 018,
  \href{http://dx.doi.org/10.1007/JHEP07(2011)018}{\doi{10.1007/JHEP07(2011)018}},
\href{http://www.arXiv.org/abs/1105.0020}{\texttt{arXiv:1105.0020}}.
%%CITATION = ARXIV:1105.0020;%%.

\bibitem{Campbell:1999ah}
\hrefCMSnoop {}{J.~M. Campbell and R.~K. Ellis, ``Update on vector boson pair
  production at hadron colliders'',} \textit{ Phys. Rev. D} \textbf{ 60} (1999)
  113006,
  \href{http://dx.doi.org/10.1103/PhysRevD.60.113006}{\doi{10.1103/PhysRevD.60.113006}},
\href{http://www.arXiv.org/abs/hep-ph/9905386}{\texttt{arXiv:hep-ph/9905386}}.
%%CITATION = HEP-PH/9905386;%%.

\bibitem{Ball:2012cx}
\hrefCMSnoop {}{{NNPDF} Collaboration, ``{Parton distributions with LHC
  data}'',} \textit{ Nucl. Phys. B} \textbf{ 867} (2013) 244,
  \href{http://dx.doi.org/10.1016/j.nuclphysb.2012.10.003}{\doi{10.1016/j.nuclphysb.2012.10.003}},
\href{http://www.arXiv.org/abs/1207.1303}{\texttt{arXiv:1207.1303}}.
%%CITATION = ARXIV:1207.1303;%%.

\bibitem{Alwall:2014hca}
J.~Alwall\hrefCMSnoop {}{ {et~al.}, ``The automated computation of tree-level
  and next-to-leading order differential cross sections, and their matching to
  parton shower simulations'',} \textit{ JHEP} \textbf{ 07} (2014) 079,
  \href{http://dx.doi.org/10.1007/JHEP07(2014)079}{\doi{10.1007/JHEP07(2014)079}},
\href{http://www.arXiv.org/abs/1405.0301}{\texttt{arXiv:1405.0301}}.
%%CITATION = ARXIV:1405.0301;%%.

\bibitem{Ahmed:2017}
T.~Ahmed\hrefCMSnoop {}{ {et~al.}, ``{NNLO} {QCD} corrections to the
  {D}rell--{Y}an cross section in models of {TeV}-scale gravity'',} \textit{
  Eur. Phys. J. C} \textbf{ 77} (2017) 22,
  \href{http://dx.doi.org/10.1140/epjc/s10052-016-4587-6}{\doi{10.1140/epjc/s10052-016-4587-6}},
  \href{http://www.arXiv.org/abs/1606.08454}{\texttt{arXiv:1606.08454}}.

\bibitem{CMS-PAS-LUM-17-001}
\href {https://cds.cern.ch/record/2257069}{{CMS Collaboration}, ``{CMS}
  luminosity measurements for the 2016 data taking period'',} CMS Physics
  Analysis Summary CMS-PAS-LUM-17-001, CERN, 2017.

\bibitem{CMS-NOTE-2011-005}
\href {https://cds.cern.ch/record/1379837}{{The ATLAS Collaboration, The CMS
  Collaboration, The LHC Higgs Combination Group}, ``Procedure for the {LHC}
  {Higgs} boson search combination in {Summer} 2011'',} Technical Report
  CMS-NOTE-2011-005, ATL-PHYS-PUB-2011-11, CERN, 2011.

\bibitem{RooStats}
L.~Moneta\href
  {http://pos.sissa.it/archive/conferences/093/057/ACAT2010_057.pdf}{ {et~al.},
  ``The {RooStats} project'',} in \textit{ $13^\text{th}$ International
  Workshop on Advanced Computing and Analysis Techniques in Physics Research
  (ACAT2010)}.
\newblock SISSA, 2010.
\newblock \href{http://www.arXiv.org/abs/1009.1003}{\texttt{arXiv:1009.1003}}.
\newblock
{PoS} (ACAT2010) 057.
%%CITATION = ARXIV:1009.1003;%%.

\bibitem{theta}
\href {http://www-ekp.physik.uni-karlsruhe.de/\~ott/theta/theta-auto}{J.~Ott,
  ``\textsc{Theta}---{A} framework for template-based modeling and
  inference'',} 2010.
\newblock \url {http://www-ekp.physik.uni-karlsruhe.de/\~ott/theta/theta-auto}.

\end{thebibliography}\endgroup
\cleardoublepage \appendix\section{The CMS Collaboration \label{app:collab}}\begin{sloppypar}\hyphenpenalty=5000\widowpenalty=500\clubpenalty=5000\vskip\cmsinstskip
\textbf{Yerevan Physics Institute, Yerevan, Armenia}\\*[0pt]
A.M.~Sirunyan, A.~Tumasyan
\vskip\cmsinstskip
\textbf{Institut f\"{u}r Hochenergiephysik, Wien, Austria}\\*[0pt]
W.~Adam, F.~Ambrogi, E.~Asilar, T.~Bergauer, J.~Brandstetter, M.~Dragicevic, J.~Er\"{o}, A.~Escalante~Del~Valle, M.~Flechl, R.~Fr\"{u}hwirth\cmsAuthorMark{1}, V.M.~Ghete, J.~Hrubec, M.~Jeitler\cmsAuthorMark{1}, N.~Krammer, I.~Kr\"{a}tschmer, D.~Liko, T.~Madlener, I.~Mikulec, N.~Rad, H.~Rohringer, J.~Schieck\cmsAuthorMark{1}, R.~Sch\"{o}fbeck, M.~Spanring, D.~Spitzbart, W.~Waltenberger, J.~Wittmann, C.-E.~Wulz\cmsAuthorMark{1}, M.~Zarucki
% ...
\vskip\cmsinstskip
% ...
1:  Also at Vienna University of Technology, Vienna, Austria\\
% ...
\end{sloppypar}
\end{document}